\PassOptionsToPackage{unicode}{hyperref}
\PassOptionsToPackage{hyphens}{url}
\PassOptionsToPackage{dvipsnames,svgnames,x11names}{xcolor}
\documentclass[
  11pt,
]{article}
\usepackage{amsmath,amssymb}
\usepackage{iftex}
\ifPDFTeX
  \usepackage[T1]{fontenc}
  \usepackage[utf8]{inputenc}
  \usepackage{textcomp} 
\else 
  \usepackage{unicode-math} 
  \defaultfontfeatures{Scale=MatchLowercase}
  \defaultfontfeatures[\rmfamily]{Ligatures=TeX,Scale=1}
\fi
\ifPDFTeX\else
\fi
\IfFileExists{upquote.sty}{\usepackage{upquote}}{}
\IfFileExists{microtype.sty}{
}{}
\makeatletter
\@ifundefined{KOMAClassName}{
  \IfFileExists{parskip.sty}{%
    \usepackage{parskip}
  }{
    \setlength{\parindent}{0pt}
    \setlength{\parskip}{6pt plus 2pt minus 1pt}}
}{
  \KOMAoptions{parskip=half}}
\makeatother
\usepackage{xcolor}
\usepackage[margin=1in]{geometry}
\usepackage{color}
\usepackage{fancyvrb}

\DefineVerbatimEnvironment{Highlighting}{Verbatim}{commandchars=\\\{\}}
\newenvironment{Shaded}{}{}

\newcommand{\BuiltInTok}[1]{\textcolor[rgb]{0.00,0.50,0.00}{#1}}

\newcommand{\CommentTok}[1]{\textcolor[rgb]{0.38,0.63,0.69}{\textit{#1}}}

\newcommand{\DataTypeTok}[1]{\textcolor[rgb]{0.56,0.13,0.00}{#1}}
\newcommand{\DecValTok}[1]{\textcolor[rgb]{0.25,0.63,0.44}{#1}}

\newcommand{\ErrorTok}[1]{\textcolor[rgb]{1.00,0.00,0.00}{\textbf{#1}}}

\newcommand{\FunctionTok}[1]{\textcolor[rgb]{0.02,0.16,0.49}{#1}}

\newcommand{\KeywordTok}[1]{\textcolor[rgb]{0.00,0.44,0.13}{\textbf{#1}}}
\newcommand{\NormalTok}[1]{#1}
\newcommand{\OperatorTok}[1]{\textcolor[rgb]{0.40,0.40,0.40}{#1}}
\newcommand{\OtherTok}[1]{\textcolor[rgb]{0.00,0.44,0.13}{#1}}

\newcommand{\StringTok}[1]{\textcolor[rgb]{0.25,0.44,0.63}{#1}}

\usepackage{longtable,booktabs,array}
\usepackage{calc} 
\usepackage{etoolbox}
\makeatletter
\patchcmd\longtable{\par}{\if@noskipsec\mbox{}\fi\par}{}{}
\makeatother
\IfFileExists{footnotehyper.sty}{\usepackage{footnotehyper}}{\usepackage{footnote}}
\makesavenoteenv{longtable}
\providecommand{\tightlist}{%
  \setlength{\itemsep}{0pt}\setlength{\parskip}{0pt}}
\ifLuaTeX
  \usepackage{selnolig}  
\fi
\usepackage[]{natbib}
\IfFileExists{bookmark.sty}{\usepackage{bookmark}}{\usepackage{hyperref}}
\IfFileExists{xurl.sty}{\usepackage{xurl}}{} 
\hypersetup{
  pdftitle={Agentic Settlement Protocol: An Application Profile for Refundable, Delayed-Fulfilment Agent Commerce on Stablecoin Rails},
  pdfauthor={Behnam (Beny) Mohammadkhani --- XDC AI --- beny@xinfin.org --- X: @zerobeny; Atul Khekade --- XDC AI --- atul@xinfin.org --- X: @atulkhekade; Ritesh Kakkad --- XDC AI --- rik@xinfin.org --- X: @riteshkakkad},
  colorlinks=true,
  linkcolor={Maroon},
  filecolor={Maroon},
  citecolor={Blue},
  urlcolor={Blue},
  pdfcreator={LaTeX via pandoc}}

\title{Agentic Settlement Protocol: An Application Profile for Refundable, Delayed-Fulfilment Agent Commerce on Stablecoin Rails}
\author{Behnam (Beny) Mohammadkhani --- XDC AI --- beny@xinfin.org --- X: @zerobeny \and Atul Khekade --- XDC AI --- atul@xinfin.org --- X: @atulkhekade \and Ritesh Kakkad --- XDC AI --- rik@xinfin.org --- X: @riteshkakkad}
\date{Draft v0.6 --- 31 August 2026}

\begin{document}
\maketitle
\begin{abstract}
Autonomous agents can already pay per request: HTTP-native protocols such as x402 let an agent sign a stablecoin authorization and receive a resource in the same round trip. That model is atomic, instant and final, which is what metered access needs and what commerce cannot tolerate. A purchase made on a person's behalf --- a service appointment, a physical order, a freelance deliverable, a flight, a hotel night, an invoice --- is large, is frequently cancelled, and should not become the seller's money until the seller has actually delivered. We present the Agentic Settlement Protocol (ASP), an application profile that specifies how a buyer agent, a seller, an operator and an on-chain authorize-and-capture escrow coordinate a quote, a time-bounded hold, fulfilment, capture and refund for any online business whose booking, order, scheduling or invoicing system can expose, or be adapted to expose, a hold, a commit, a cancel and a refund computation. ASP treats all such systems uniformly as \emph{fulfilment engines}, and defines the lifecycle once against a registry of fulfilment classes rather than per vertical. ASP does not introduce a new escrow primitive; it builds on the auth-capture escrow pattern already standardised by the Commerce Payments Protocol and adopted by x402 refund extensions. Its contributions are a three-deadline hold model --- an issuance deadline, an escrow expiry and the engine's own inventory expiry, separated by an explicit submission-inclusion-finality margin --- under which no inventory is ever issued against funds the buyer can reclaim; a fulfilment-verification ladder that states explicitly which party is trusted to trigger capture, what its assertion does and does not prove, and what challenge window applies; fare-rule-driven partial refunds with a specified refund-liquidity order and per-seller exposure controls that bound the operator's credit risk; distributor revenue share taken atomically at capture with a condition under which self-dealing is provably unprofitable; a single-currency-per-charge invariant that removes swaps, oracles and slippage from the settlement path; a fulfilment-class registry with worked mappings for order-management, service-scheduling, invoicing and travel systems that leaves the seller's core engine unchanged and integrates through a connector; and a normative interface specification (x402 scheme, vault interface, operator and connector APIs, registries, conformance levels) intended to let independent implementations interoperate. The design originated in a review of how travel agencies could join agentic settlement layers, and travel remains the running example because it forces every part of the design to exist; nothing in the protocol is specific to it. We state the trust model candidly, including where the operator is trusted, give informal arguments for five safety properties, and describe an instantiation on the XDC Network. ASP is presented as a design and specification together with a narrow reference-implementation scope and a fault-injection evaluation plan; measured results are left to follow-up work.
\end{abstract}

\hypertarget{introduction}{%
\section{Introduction}\label{introduction}}

A payment standard already exists for AI agents, and it solves the wrong half of commerce. x402 revived the HTTP 402 status code so that an agent can request a resource, receive a machine-readable price, sign a stablecoin authorization and receive the response in a single exchange \citep{x402}. The protocol has since moved to the Linux Foundation, gained multiple settlement schemes, and reports hundreds of millions of payments \citep{x402lf, x402schemes}. It is the right design for data, tools and inference: one-shot, instant, irreversible, and small.

The moment an agent tries to buy something real, that shape breaks. A plumber's appointment, a physical order from an online store, a design deliverable from a freelancer, a seat on a flight or a night in a hotel all share three properties that a metered API call does not. The amount is large relative to the buyer's per-call budget. The purchase is refundable, and in many service businesses cancellation is routine rather than exceptional. And, critically, the money should not become the seller's until the seller has delivered: the technician turned up, the parcel shipped, the file was accepted, the ticket issued. A single atomic transfer has no room to hold funds, wait for fulfilment, and then either settle or return them.

The traditional answer is the card network's authorize--capture--refund lifecycle executed by a processor. That lifecycle is the correct abstraction, but its implementation is custodial, fiat-bound and closed to any seller without a processor relationship. In 2025 Coinbase and Shopify brought the same lifecycle on-chain in the Commerce Payments Protocol (CPP), a non-custodial escrow contract with authorize, capture, charge, void, reclaim and refund operations, pluggable token collectors, and time-bounded holds that the buyer can unilaterally reclaim on expiry \citep{cpp, cppblog}. Refund extensions to x402 now wire that escrow into the 402 handshake \citep{x402r}. The on-chain primitive therefore exists and is audited.

What does not exist is a specification of how that primitive should be \emph{driven} when fulfilment is decided by an off-chain system that has its own clock, its own hold semantics and its own refund rules. Every online business already runs one: an order-management system with a fulfilment deadline, a scheduling system with an appointment slot that lapses if unconfirmed, an invoicing system with a due date, a booking engine with a ticketing time limit. If the escrow hold and the business's own hold are not coordinated, one of two failures occurs: money is held after the slot or stock has been released, or the slot is held after the money has gone back to the buyer. Neither CPP nor the x402 schemes say anything about this, because they are payment-layer specifications and the problem sits one layer up.

This paper specifies that layer. The Agentic Settlement Protocol (ASP) is an application profile over an auth-capture escrow for purchases from any online business whose fulfilment system can expose, or be adapted to expose, the four operations below; that covers most businesses that sell online, but it is an interface requirement, not a universal claim. We abstract the business's system as a \emph{fulfilment engine} exposing a hold, a commit, a cancel and a refund computation, define the settlement lifecycle once against that abstraction, and instantiate it per business type through a registry of fulfilment classes. Travel is the running example throughout, for two reasons. The first is historical: this work began as a review of how travel agencies could participate in agentic AI settlement layers, and the design was worked out against their booking engines, ticketing time limits and fare rules before it became clear that nothing in it was specific to travel. The second is that travel is the hardest instance --- the holds are short, refunds are rule-driven and partial, and the engines are decades old --- so a design that survives it survives easier verticals. A reader should substitute their own business type at every occurrence. We make the following contributions.

\begin{enumerate}
\def\labelenumi{\arabic{enumi}.}
\tightlist
\item
  \textbf{A three-deadline hold model} (Section 4.5) that separates the issuance deadline the connector enforces, the escrow expiry, and the engine's inventory expiry by an explicit margin covering transaction submission, inclusion and finality, together with a proposition that no inventory is issued against funds the buyer can reclaim. The model does not assume that two independent systems cancel each other atomically.
\item
  \textbf{An explicit fulfilment-verification ladder} (Section 4.7) that names, for each class of good, which party's assertion triggers capture, what evidence it carries, and what challenge window applies. This replaces the phrase ``on fulfilment'' that most agent-payment designs leave undefined.
\item
  \textbf{A refund and exposure model} (Section 4.6) in which the off-chain engine's fare or rate rules are authoritative for the refundable amount, with a specified refund-liquidity order, a refund window, and per-seller exposure controls (rolling reserve, settlement delay, outstanding-refundable caps) that make the operator's refund backstop a bounded credit exposure rather than an open-ended guarantee.
\item
  \textbf{Revenue share at capture} (Section 4.8), with a proposition giving the condition under which a distributor cannot profit by generating its own referred volume.
\item
  \textbf{A single-currency invariant} (Section 4.9), a fulfilment-class registry with worked mappings for order-management, service-scheduling, invoicing and travel systems (Section 6) that leaves the seller's core engine unchanged, and a deliberately narrow reference-implementation scope with a fault-injection evaluation plan (Sections 7 and 9).
\item
  \textbf{A normative interface specification} (Section 4.11 and Appendices C--F): the x402 \texttt{asp} scheme, the vault's Solidity interface with events and errors, the operator and connector APIs, the registry and webhook schemas, and conformance levels. The intent is that two independent implementations built from the appendices interoperate.
\end{enumerate}

We state the trust model in Section 3 without euphemism: the operator is trusted for the capture decision to the extent set by the verification rung, and the ``instant reserve'' settlement mode is custodial by construction. Section 5 gives informal arguments for five safety properties. Section 7 describes the instantiation on the XDC Network. Section 8 situates ASP against both industry protocols and the recent academic literature on agent payments, and Section 9 lists limitations and open questions. ASP is a design contribution; we have not yet measured a deployed vault, and we say so, but Section 9 specifies exactly what will be measured.

\hypertarget{background}{%
\section{Background}\label{background}}

\hypertarget{metered-agent-payments-x402}{%
\subsection{Metered agent payments: x402}\label{metered-agent-payments-x402}}

x402 defines an HTTP handshake in which a server answers an unpaid request with \texttt{402\ Payment\ Required} and a list of acceptable payment requirements; the client signs a payment payload and retries; a facilitator verifies the payload and submits settlement on the server's behalf \citep{x402}. Version 2 moved the payload into standard headers, adopted CAIP chain identifiers and made networks and facilitators pluggable \citep{x402lf}. Three schemes are specified: \texttt{exact}, where the buyer authorises the advertised amount; \texttt{upto}, where the buyer authorises a maximum and the seller charges actual usage; and \texttt{batch-settlement}, where the buyer deposits into an on-chain escrow once, signs an off-chain voucher per request, and the seller redeems many vouchers in one transaction \citep{x402schemes}. Cloudflare has separately proposed a \texttt{deferred} scheme that decouples the cryptographic handshake from settlement entirely \citep{cloudflare402}.

All three shipped schemes share one property: the seller's entitlement to funds is decided by the buyer's signature and the seller's own accounting. There is no notion of a hold that lapses if the seller fails to deliver, and \texttt{exact} is a push payment that cannot be reversed after execution \citep{x402lf}. Refunds, where they exist, are a separate remittance.

\hypertarget{on-chain-authorize-and-capture-escrow-cpp-and-x402r}{%
\subsection{On-chain authorize-and-capture escrow: CPP and x402r}\label{on-chain-authorize-and-capture-escrow-cpp-and-x402r}}

The Commerce Payments Protocol places a non-custodial escrow contract between payer and receiver and drives it through an \emph{operator} address. Six operations are defined: \texttt{authorize} (funds move from payer to escrow under a time-bounded hold), \texttt{capture} (escrow pays the receiver, less a fee within a pre-declared range), \texttt{charge} (authorize and capture in one call), \texttt{void} (operator cancels an authorization), \texttt{reclaim} (payer unilaterally recovers an expired authorization) and \texttt{refund} (funds return to payer after capture, sourced from a \emph{refund collector} that may draw on the receiver, another merchant address, or the operator) \citep{cpp, cppblog}. Token collection is modular over ERC-3009, Permit2, allowances and spend permissions \citep{erc3009, permit2, eip2612}. Refunds must occur before a refund expiry after which the capture is final. x402r packages this escrow as a \texttt{commerce} scheme inside the x402 handshake, with dispute resolution pluggable per arbiter \citep{x402r}.

CPP is the primitive ASP builds on. Where this paper says ``the escrow'', a CPP-compatible contract satisfies every requirement; where ASP names its own vault, it is to fix a minimal interface (Section 4.2) and the XDC deployment (Section 7), not to claim a new mechanism.

\hypertarget{agent-authorization-and-checkout-ap2-and-acp}{%
\subsection{Agent authorization and checkout: AP2 and ACP}\label{agent-authorization-and-checkout-ap2-and-acp}}

Google's Agent Payments Protocol (AP2) standardises the consent chain: a signed \emph{intent mandate} records the user's goal and spend cap, a \emph{cart mandate} records the agent's specific selection, and a \emph{payment mandate} records the charge \citep{ap2}. Lan et al.~show that AP2 implementations are exposed to replay and context-binding failures unless mandates are bound to the runtime context that produced them \citep{ap2replay}. The Agentic Commerce Protocol (ACP) from Stripe and OpenAI standardises the checkout messages between an agent and a merchant, with a processor performing authorization, capture and refund off-chain \citep{acp}. Neither protocol specifies on-chain settlement; both are the layers ASP expects to sit beneath.

\hypertarget{fulfilment-lifecycles-in-online-businesses}{%
\subsection{Fulfilment lifecycles in online businesses}\label{fulfilment-lifecycles-in-online-businesses}}

Almost every system that sells online already implements a hold-then-commit lifecycle, whether or not it calls it that. An order-management system (Shopify, WooCommerce, Salesforce Commerce, a custom OMS) creates an order, reserves stock, and moves to \emph{fulfilled} when a shipment or pickup is confirmed; unreserved stock is released after a timeout and refunds follow a returns policy. A service-scheduling system (Calendly-style, field-service platforms, marketplaces for trades and professionals) tentatively holds a slot, confirms it, and marks the job complete; cancellation fees follow a notice-period rule. An invoicing or procurement system issues an invoice with a due date and marks it paid or disputed. A freelance or gig marketplace opens a task, accepts a deliverable, and releases payment on acceptance. In travel, a global distribution system creates a passenger name record with a \emph{ticketing time limit} after which the seat is released; IATA's NDC standard has \texttt{OrderCreate} with an order time limit and \texttt{OrderChange} for modification and refund \citep{ndc}; hotels expose an \emph{option} with an option date, then a confirmed reservation, then cancellation under rate rules.

The common shape is: a hold with a time limit, a commit that produces a reference the buyer can later check, a cancel that releases the hold, and a refund computation the engine owns. ASP calls any such system a \emph{fulfilment engine} and requires exactly those four operations of it (Section 4). Refund amounts are computed by the engine from its own rules and are rarely a flat window, in every one of these verticals.

\hypertarget{system-and-trust-model}{%
\section{System and Trust Model}\label{system-and-trust-model}}

\hypertarget{parties}{%
\subsection{Parties}\label{parties}}

\begin{itemize}
\tightlist
\item
  \textbf{Principal.} The human or organisation on whose behalf the purchase is made. Establishes the agent's spend cap (an AP2 intent mandate or equivalent).
\item
  \textbf{Buyer agent.} Software that selects the offer and signs the authorization. Holds funds in a smart-contract wallet with an on-chain spending limit (Section 7).
\item
  \textbf{Seller.} Owns the inventory and operates, or subscribes to, a booking or order engine. Receives capture proceeds to a payout vault.
\item
  \textbf{Operator.} Runs the protocol: publishes offers, relays signed authorizations to the escrow, observes fulfilment evidence, and calls \texttt{capture} or \texttt{void}. In the reference deployment the operator is XDC AI.
\item
  \textbf{Distributor (optional).} A party that referred the buyer agent and earns a configured share of the fee.
\item
  \textbf{Escrow.} An on-chain contract holding each charge's funds between authorization and capture.
\item
  \textbf{Fulfilment engine.} The seller's order-management, scheduling, invoicing, marketplace or booking system, exposed through an API the connector can drive. Authoritative for whether fulfilment occurred and for the refundable amount.
\item
  \textbf{Arbiter (optional).} A party or contract that can resolve a challenge during a challenge window (Section 4.7).
\end{itemize}

\hypertarget{assumptions}{%
\subsection{Assumptions}\label{assumptions}}

\begin{itemize}
\tightlist
\item
  \textbf{A1 (Timing).} Three delays are distinguished. \(T_s\) bounds the operator's time from deciding to capture to broadcasting the transaction (relayer, nonce management, RPC availability); \(T_i\) bounds the time from broadcast to inclusion in a block (congestion, fee policy); \(T_f\) bounds the time from inclusion to deterministic finality. Block timestamps drift from wall-clock time by at most \(\delta\). Only \(T_f\) is a property of the chain; \(T_s\) and \(T_i\) are operational bounds the operator chooses and is responsible for. Define the capture margin \(M = T_s + T_i + T_f + \delta\).
\item
  \textbf{A2 (Stablecoin).} The settlement token is a fiat-referenced stablecoin supporting at least one gasless authorization method (ERC-3009, EIP-2612 or Permit2). Depeg risk is handled by a pause (Section 4.10), not by the protocol.
\item
  \textbf{A3 (Signatures).} EIP-712 typed signatures are unforgeable; the buyer agent's key is under the principal's spend policy \citep{eip712}.
\item
  \textbf{A4 (Operator liveness).} The operator eventually submits transactions; if it does not, the buyer's recovery path does not depend on it (Property P1).
\item
  \textbf{A5 (Engine honesty).} The fulfilment engine reports issuance, cancellation and refundable amounts correctly to the connector. ASP treats the engine as an oracle for its own inventory.
\item
  \textbf{A6 (Operator honesty, bounded).} The operator may be malicious with respect to \emph{timing} (delaying capture or void) but is trusted not to fabricate fulfilment evidence at the verification rung in force for the charge. Section 4.7 sets out how this trust is reduced.
\end{itemize}

\hypertarget{what-is-and-is-not-trust-minimised}{%
\subsection{What is and is not trust-minimised}\label{what-is-and-is-not-trust-minimised}}

Two claims in earlier descriptions of this design need qualifying. ``Non-custodial'' holds in the sense that funds in flight sit in a contract with deterministic rules and cannot be moved by the operator to an arbitrary address; it does not mean the operator has no discretion. The operator decides when to capture, and if it captures without fulfilment the buyer's remedy is the refund path and the challenge window, not the escrow itself. This is the same trust position as CPP's operator \citep{cppblog}. Second, the \emph{instant reserve} settlement mode (Section 4.10), in which the seller is paid at authorization and refunds are drawn from a rolling reserve, is custodial by construction: the reserve is operator-controlled. It is offered as a vetted-seller option, not as the default, and is labelled as such.

\hypertarget{threats-considered}{%
\subsection{Threats considered}\label{threats-considered}}

Double-charging a booking on retry; stranded funds when a hold is never captured; capture without fulfilment; refund default when a seller's vault is empty; a compromised buyer agent draining the principal; a seller hook redirecting settlement funds; a distributor generating its own referred volume to farm revenue share; and a stablecoin depeg mid-hold. Threats \emph{not} in scope: compromise of the fulfilment engine itself, chain-level censorship beyond \(T_f\), and privacy of on-chain charge metadata (Section 9).

\hypertarget{protocol}{%
\section{Protocol}\label{protocol}}

\hypertarget{objects}{%
\subsection{Objects}\label{objects}}

An \textbf{Offer} is a priced, signed quotation with a short expiry. It is the document that fixes every settlement term the safety rules depend on, so those terms are fields of the Offer rather than operator-side configuration:

\begin{verbatim}
Offer {
  sellerId, description, amount, currency,
  fulfilmentClass,      // registry entry, Section 6.1
  engineRef,            // handle to the provisional engine hold
  engineExpiry,         // t_e as reported by the engine (or policy default)
  issueDeadline,        // t_issue, Section 4.5
  holdExpiresAt,        // t_h, Section 4.5
  verificationRung,     // D | C | A | S  (Section 4.7)
  challengeWindow,      // W_c
  refundWindow,         // W_r
  refundPolicyRef,      // pointer to the engine's refund rules
  expiresAt,            // offer validity
  signer, signerAuthority, signature   // Section 4.3
}
offerId = keccak256(EIP-712 encoding of Offer)
\end{verbatim}

Because \texttt{offerId} is the hash of the whole Offer, a buyer authorization that names \texttt{offerId} commits the buyer to all of these terms, and no party can alter \(t_{\text{issue}}\), \(t_h\), the rung or the windows after authorization without producing a different \texttt{offerId} that the escrow will not match.

A \textbf{Charge} is the payment intent and its lifecycle record:

\begin{verbatim}
Charge {
  chargeId, offerId, buyer, sellerVault, distributor,
  amount, currency,                 // one currency, no swap
  feeBps, feeAmount, distributorShareBps,
  status,                           // Section 4.2
  issueDeadline, holdExpiresAt,     // on-chain, copied from the Offer
  idempotencyKey,                   // = chargeId
  capturedAmount, refundedAmount,   // refundedAmount <= capturedAmount
  captureEvidence, refundRecords[]
}
\end{verbatim}

A \textbf{FulfilmentAttestation} is the evidence that moves a charge from \texttt{authorized} to \texttt{captured}. It is a connector-signed \emph{normalised receipt} whose hash is committed on-chain:

\begin{verbatim}
FulfilmentReceipt {                // signed off-chain by the connector
  chargeId, supplierId, orderId,   // engine order / booking / job / invoice id
  fulfilmentRef,                   // ticket no., tracking no., completion id...
  subjectHash,                     // keccak(normalised buyer/recipient data)
  issuedAt, rung
}
FulfilmentAttestation {            // passed to capture()
  chargeId, receiptHash,           // keccak(FulfilmentReceipt)
  issuedAt, attestor, attestorSig
}
\end{verbatim}

The receipt gives unambiguous attribution: the connector, operated by the operator, asserts that the engine reported issuance with these references. It does not prove that the supplier issued anything; Section 4.7 is explicit about that.

\texttt{chargeId} is the EIP-712 hash of the buyer's authorization struct, so it doubles as the idempotency key at every layer.

\hypertarget{state-machine-and-escrow-interface}{%
\subsection{State machine and escrow interface}\label{state-machine-and-escrow-interface}}

\begin{verbatim}
              deposit                 capture
  created ------------> authorized -------------> captured <--+ refund(partial):
     |                      |                        |        |   refundedAmount +=,
     | expire/reject        | void | expire          |        +-- stays captured
     v                      v                        | refund with cumulative
  cancelled              reclaimed                   |   == capturedAmount
                                                     v
                                                 refunded
\end{verbatim}

Transitions and who may trigger them:

\begin{longtable}[]{@{}
  >{\raggedright\arraybackslash}p{(\columnwidth - 6\tabcolsep) * \real{0.1444}}
  >{\raggedright\arraybackslash}p{(\columnwidth - 6\tabcolsep) * \real{0.1333}}
  >{\raggedright\arraybackslash}p{(\columnwidth - 6\tabcolsep) * \real{0.3778}}
  >{\raggedright\arraybackslash}p{(\columnwidth - 6\tabcolsep) * \real{0.3444}}@{}}
\toprule\noalign{}
\begin{minipage}[b]{\linewidth}\raggedright
From
\end{minipage} & \begin{minipage}[b]{\linewidth}\raggedright
To
\end{minipage} & \begin{minipage}[b]{\linewidth}\raggedright
Trigger
\end{minipage} & \begin{minipage}[b]{\linewidth}\raggedright
Caller
\end{minipage} \\
\midrule\noalign{}
\endhead
\bottomrule\noalign{}
\endlastfoot
created & authorized & \texttt{deposit(auth)} & operator (relays buyer sig) \\
created & cancelled & offer expired / buyer declines & anyone \\
authorized & captured & \texttt{capture(chargeId,\ attestation)} & operator \\
authorized & reclaimed & \texttt{void(chargeId)} & operator \\
authorized & reclaimed & \texttt{reclaim(chargeId)} after expiry & buyer, or anyone on behalf \\
captured & captured & \texttt{refund(chargeId,\ amount)}, cumulative \(<\) captured & operator, within refund window \\
captured & refunded & \texttt{refund(chargeId,\ amount)}, cumulative \(=\) captured & operator, within refund window \\
\end{longtable}

The escrow interface ASP requires is a strict subset of CPP: \texttt{deposit} (CPP \texttt{authorize}), \texttt{capture}, \texttt{void}, \texttt{reclaim} and \texttt{refund}. Every call is keyed by \texttt{chargeId} and is idempotent: a repeated \texttt{deposit} with the same key is a no-op, a repeated \texttt{capture} returns the existing settlement, and \texttt{refund} accumulates \texttt{refundedAmount}, which cannot exceed \texttt{capturedAmount}; the charge remains \texttt{captured} while further refund is possible and becomes \texttt{refunded} only when the two are equal. On-chain, \texttt{deposit} also records \texttt{issueDeadline} from the authorization, and \texttt{capture} reverts if the attestation's \texttt{issuedAt} exceeds it, so rule H3 (Section 4.5) is enforced by the vault rather than by operator discipline.

\emph{Terminology note.} An earlier draft used \texttt{release} for the return of funds to the buyer. Because \texttt{release} conventionally means paying the seller, this draft uses \texttt{void} (operator-initiated) and \texttt{reclaim} (buyer-initiated after expiry) throughout.

\hypertarget{who-signs-the-offer}{%
\subsection{Who signs the Offer}\label{who-signs-the-offer}}

An Offer binds the seller commercially, so its signer matters. Three models are defined, recorded in the Offer's \texttt{signerAuthority} field:

\begin{itemize}
\tightlist
\item
  \textbf{Direct.} The seller's own key signs each Offer. Appropriate for direct integrations with low quote volume.
\item
  \textbf{Delegated.} The seller registers a connector key with a scope (fulfilment classes, maximum amount, validity period) in the operator's seller registry, and the connector signs Offers under that key. The seller can revoke the delegation at any time; Offers signed after revocation are invalid. This is the model for platform connectors, where thousands of sellers on one commerce or scheduling platform cannot sign machine-generated quotes individually.
\item
  \textbf{Operator-signed under mandate.} The seller signs a standing mandate authorising the operator to issue Offers within stated limits; the operator signs Offers and includes the mandate reference. Functionally similar to delegated, but the key is the operator's and the trust boundary is correspondingly wider.
\end{itemize}

In the delegated and mandate models the party holding the signing key can bind the seller to prices and terms within scope. That is a real trust grant, and the scope limits are what bound it; the registry makes the grant and its limits inspectable. ASP-Lite uses the delegated model.

\hypertarget{message-flow}{%
\subsection{Message flow}\label{message-flow}}

The sequencing question every implementation must answer is when the engine's hold is created, because \(t_e\) has to be known before \(t_h\) and \(t_{\text{issue}}\) can be derived, and those have to be in the Offer before the buyer signs. ASP therefore places the engine hold \emph{before} the Offer:

\begin{enumerate}
\def\labelenumi{\arabic{enumi}.}
\tightlist
\item
  \textbf{Provisional hold and quote.} The buyer agent requests a price. The connector asks the fulfilment engine for a provisional hold (reserve stock, tentatively hold the slot, create an unticketed PNR); the engine returns \texttt{engineRef} and its expiry \(t_e\). The connector derives \(t_{\text{issue}}\) and \(t_h\) from \(t_e\) under rule H1 and the operator, or the delegated signer, issues the signed \texttt{Offer} with a short \texttt{expiresAt}. Price and every timing term are locked before the buyer commits.
\item
  \textbf{Authorize.} The buyer agent signs an EIP-712 \texttt{ChargeAuthorization} over \texttt{(offerId,\ buyer,\ token,\ amount,\ issueDeadline,\ holdExpiresAt,\ nonce)} (Appendix A) using the token's gasless method. The operator relays it; the escrow verifies that the authorization's timing fields match the Offer's (it recomputes \texttt{offerId} from the terms the operator supplies alongside), pulls funds, and records \texttt{issueDeadline} and \texttt{holdExpiresAt}. The charge is \texttt{authorized}; the buyer has paid nothing yet.
\item
  \textbf{Decline or lapse.} If the buyer does not authorize before \texttt{expiresAt}, or the deposit fails, the connector releases the provisional engine hold. Provisional holds are therefore short-lived and their cost is bounded by \texttt{expiresAt}.
\item
  \textbf{Commit.} Before \(t_{\text{issue}}\), the connector instructs the engine to commit against \texttt{engineRef} (ship the order, confirm the appointment, accept the deliverable, issue the ticket). The engine returns its record; the connector produces a \texttt{FulfilmentAttestation}.
\item
  \textbf{Capture.} The operator calls \texttt{capture} with the attestation. The escrow checks \texttt{issuedAt} \(\le\) \texttt{issueDeadline}, pays \texttt{feeAmount} to the fee splitter (Section 4.8) and \texttt{amount\ -\ feeAmount} to the seller's vault, in the same token. The charge is \texttt{captured}.
\item
  \textbf{Void or reclaim.} If the engine cannot commit, the operator calls \texttt{void}. If the operator does neither by \texttt{holdExpiresAt}, the buyer (or any relayer) calls \texttt{reclaim}. Either way the full amount returns to the buyer.
\item
  \textbf{Refund.} After capture, a cancellation or return is submitted to the engine; the engine computes the refundable amount under its own rules; the operator calls \texttt{refund} for exactly that amount, sourced per Section 4.6.
\end{enumerate}

\textbf{Engines that cannot hold before payment.} Some systems only create a reservation once a payment is authorized. For a fulfilment class flagged \texttt{holdBeforeAuth\ =\ false} in the registry, the connector uses a conservative \emph{policy} expiry \(t_e^{*}\) recorded in the registry for that class (for example, the platform's documented minimum reservation lifetime) in place of the engine's reported value at step 1, and creates the engine hold immediately after step 2. When the engine then returns its actual \(t_e\), the connector checks H1 against it. If \(t_e \ge t_h + \delta\) the charge proceeds; if not, the connector calls \texttt{void} at once, the funds return to the buyer, and the offer is re-quoted with the now-known \(t_e\). A charge is never allowed to continue under a violated H1. The cost of this path is one wasted deposit-void round trip on the rare shortfall, which the evaluation plan measures.

\hypertarget{hold-coordination}{%
\subsection{Hold coordination}\label{hold-coordination}}

The escrow and the fulfilment engine are two independent distributed systems with independent clocks. No rule can make them expire atomically, so ASP does not try. Instead it introduces a third deadline, enforced by the connector, that sits strictly before both.

Let \(t_e\) be the engine's hold time limit (stock reservation timeout, slot hold expiry, invoice due date, ticketing time limit) in wall-clock time; \(t_h\) the escrow's \texttt{holdExpiresAt} in chain time; and \(t_{\text{issue}}\) the \textbf{issuance deadline}, the latest time at which the connector will instruct the engine to issue against this charge.

\textbf{Rule H1 (ordering).} \(t_{\text{issue}} + M < t_h \le t_e - \delta\), with \(M = T_s + T_i + T_f + \delta\) from A1.

\textbf{Rule H2 (issuance gate).} The connector instructs issuance only if the current time is before \(t_{\text{issue}}\) and the charge is \texttt{authorized}. After \(t_{\text{issue}}\) the connector refuses to issue even if the engine's reservation is still alive.

\textbf{Rule H3 (capture gate).} \texttt{capture} succeeds only for an attestation with \texttt{issuedAt} \(\le t_{\text{issue}}\); the vault enforces this against the \texttt{issueDeadline} recorded at deposit. The operator submits within \(T_s\) of receiving the attestation.

\textbf{Rule H4 (cleanup).} When \(t_h\) passes without capture, the connector cancels the engine reservation promptly; when the engine cancels first, the connector calls \texttt{void} promptly. Promptness here is a liveness goal, not a safety requirement.

\textbf{Proposition 1 (no issuance against reclaimable funds).} Under A1 and H1--H3, no inventory is issued against a charge whose funds are, or will become before capture lands, reclaimable by the buyer.

\emph{Argument.} Issuance occurs only before \(t_{\text{issue}}\) (H2), and a capture whose attestation post-dates \(t_{\text{issue}}\) reverts on-chain (H3). Any capture that lands therefore has \texttt{issuedAt} \(\le t_{\text{issue}}\), was broadcast within \(T_s\), included within a further \(T_i\), and final within a further \(T_f\). Allowing \(\delta\) for clock drift, the capture is final by \(t_{\text{issue}} + M\), which by H1 is strictly before \(t_h\). Hence at the moment funds become reclaimable (\(t_h\)), either the charge is already \texttt{captured}, or no issuance occurred. \(\square\)

What Proposition 1 does \emph{not} say is also important. Between \(t_h\) and \(t_e\) the engine reservation may still be technically alive while the funds are reclaimable. That window is harmless because H2 forbids issuance in it; the reservation is inventory held, not inventory sold, and H4 releases it. Earlier drafts claimed there was no such window; there is, and the design tolerates it rather than pretending it away.

\textbf{Residual risk (late capture).} If the operator violates its own \(T_s + T_i\) bound --- a relayer outage, a nonce stall, sustained congestion --- a capture broadcast before \(t_h\) may be included after it and revert, while the ticket was already issued. The buyer is made whole by \texttt{reclaim}; the seller has issued and is unpaid. Because the operator sets \(M\) and operates the relayer, ASP assigns this protocol-level loss to the operator: the operator pays the seller from its own pool (Section 4.6), the liability is counted in the operator's aggregate exposure (Proposition 3), and the event is reported as a \emph{late-capture loss} in the evaluation metrics (Section 9). The seller's remaining risk is counterparty risk against the operator's pool, which is what the exposure ceiling exists to bound. The correct mitigation is a generous \(M\), not a shorter one; the reference deployment uses \(T_s = 60\) s, \(T_i = 120\) s, \(T_f\) per Section 7 and \(\delta = 30\) s, giving \(M\) under five minutes against engine hold limits that are typically minutes for a scheduling slot, hours for a stock reservation, and hours or days for travel. Fulfilment classes whose engine hold is shorter than \(M\) cannot be served in escrow mode and must use \texttt{charge} (authorize and capture in one call, rung D or C only); the registry records this per class.

\hypertarget{refunds}{%
\subsection{Refunds}\label{refunds}}

\textbf{Pre-capture.} Void and reclaim return the full amount; there is no partial pre-capture refund because nothing has been charged.

\textbf{Post-capture.} The refundable amount is whatever the fulfilment engine computes under its own rules: a returns policy for an order, a notice-period schedule for a service appointment, a dispute outcome for an invoice, fare or rate rules for travel. ASP does not impose a flat window on top of the engine; it executes the engine's number. This choice follows from A5 and keeps the seller's back office authoritative.

\textbf{Refund window.} A captured charge carries a refund window \(W_r\) (default: the longer of the engine's cancellation horizon and 30 days) after which the capture is final and the escrow will reject \texttt{refund}. This mirrors CPP's refund expiry \citep{cppblog} and bounds the seller's contingent liability at the protocol layer. Settlement becomes final \emph{under ASP} after \(W_r\); this does not extinguish any refund, warranty or consumer-protection obligation that exists outside the protocol under the law that governs the sale, and a seller or operator that owes such a refund after \(W_r\) must pay it off-protocol. ASP bounds what the escrow will do, not what the parties owe.

\textbf{Refund liquidity order.} A refund is sourced, in order, from (i) the seller's payout vault, (ii) the seller's rolling reserve, (iii) the seller's settlement-delayed proceeds not yet released, (iv) the operator's refund pool. The operator's pool is funded from fees and is the last-resort backstop; the operator recovers advances from the seller's future captures and reserve.

\textbf{Seller exposure controls.} A backstop that is unbounded is a credit line, and a seller with large refundable volume who withdraws proceeds and becomes insolvent would convert it into the operator's loss. ASP therefore requires the operator to enforce, per seller \(s\), a set of controls that bound its exposure before any charge is captured:

\begin{itemize}
\tightlist
\item
  \emph{Outstanding refundable exposure.} \(R(s) = \sum\) captured amounts within their refund window not yet refunded. This is the quantity every other control is sized against.
\item
  \emph{Rolling reserve} \(\rho(s) \in [0,1]\): a fraction of each capture retained in a reserve vault and released after the refund window closes.
\item
  \emph{Settlement delay} \(D(s)\): captured proceeds beyond the reserve are released to the seller's withdrawable balance only after \(D(s)\), so that for young or high-cancellation sellers most refundable volume is still in escrow-adjacent custody when a refund is requested.
\item
  \emph{Exposure cap} \(E_{\max}(s)\): new captures are refused (and new offers are not issued) when \(R(s) - \text{reserve}(s) - \text{delayed}(s) > E_{\max}(s)\). The uncovered exposure is the only amount the operator's pool can be asked to absorb.
\item
  \emph{Capture limits}: per-charge and per-day maximums by risk tier.
\item
  \emph{Risk tier}: a small number of tiers (new, verified, accredited) that set default values of \(\rho\), \(D\), \(E_{\max}\) and capture limits, adjusted by observed cancellation and refund rates.
\item
  \emph{Exposure reservation.} Checks against a cap are useless if a hundred concurrent quotes can each pass them before any is recorded. Exposure is therefore \emph{reserved} at Offer issuance, not measured at capture. For seller \(s\) define \[A(s) = E_{\max}(s) - \bigl[R(s) + \text{authorized}(s) + \text{committedNotCaptured}(s) + \text{reservedOffers}(s)\bigr],\] the \emph{available} exposure. An Offer for amount \(a\) MAY be issued only if \(a \le A(s)\), and issuing it atomically adds \(a\) to \texttt{reservedOffers}. The amount moves to \texttt{authorized} on deposit, to \texttt{committedNotCaptured} on commit, to \(R(s)\) on capture, and is released on Offer expiry, void or reclaim. The reservation update MUST be atomic in the risk engine (a compare-and-set or a serialisable transaction); the on-chain vault maintains the \texttt{authorized}, \texttt{committedNotCaptured} and \(R\) terms itself, and the operator adds \texttt{reservedOffers}. Without reservation, a seller at USD 90k of a USD 100k cap could be quoted a hundred USD 1k Offers simultaneously and end at USD 190k of potential liability.
\item
  \emph{Operator aggregate liability}: the operator's pool backstops more than refunds. Define \[L = \sum_s \max\bigl(0,\, R(s) - \text{reserve}(s) - \text{delayed}(s)\bigr) \;+\; C_M \;+\; G,\] where \(C_M\) is the \emph{late-capture exposure}, the value of charges committed but not yet captured (bounded by the volume committed in any window of length \(M\), Section 4.5), and \(G\) is any other settlement guarantee the operator has extended (for example advances under instant-reserve mode). The \emph{operator exposure ceiling} requires \(L \le \lambda \cdot \text{pool}\) for a fixed \(\lambda < 1\); new offers are not issued while it is breached.
\end{itemize}

\textbf{Proposition 3 (bounded operator exposure).} If every seller's reserved-inclusive exposure is within its cap, the aggregate ceiling holds, and reservation updates are atomic, the operator's maximum loss from the simultaneous default of all sellers together with the failure of every in-flight capture is at most \(\lambda \cdot \text{pool}\).

\emph{Argument.} Refund claims beyond a seller's own reserve and delayed proceeds are limited to that seller's uncovered exposure, which is a term of \(L\); late-capture claims are limited to \(C_M\); other guarantees are \(G\). The ceiling bounds their sum. Bounding each term separately would not suffice, since all three are claims on the same pool; that is why the ceiling is on \(L\). Atomic reservation is needed because otherwise concurrent Offers could each be admitted against the same headroom and the bound would hold only for exposure already recorded, not for exposure already promised. \(\square\)

The controls are enforced off-chain by the operator except the reserve and settlement delay, which are enforced by the vault's fee splitter at capture. They are parameters of the deployment, not of the protocol; the protocol requires only that they exist and that \(R(s)\) is computable from on-chain state.

\hypertarget{fulfilment-verification-ladder}{%
\subsection{Fulfilment verification ladder}\label{fulfilment-verification-ladder}}

The phrase ``capture on fulfilment'' is only meaningful once one says who asserts fulfilment. ASP assigns each offer a rung:

\begin{longtable}[]{@{}
  >{\raggedright\arraybackslash}p{(\columnwidth - 8\tabcolsep) * \real{0.0411}}
  >{\raggedright\arraybackslash}p{(\columnwidth - 8\tabcolsep) * \real{0.1164}}
  >{\raggedright\arraybackslash}p{(\columnwidth - 8\tabcolsep) * \real{0.4795}}
  >{\raggedright\arraybackslash}p{(\columnwidth - 8\tabcolsep) * \real{0.1233}}
  >{\raggedright\arraybackslash}p{(\columnwidth - 8\tabcolsep) * \real{0.2397}}@{}}
\toprule\noalign{}
\begin{minipage}[b]{\linewidth}\raggedright
Rung
\end{minipage} & \begin{minipage}[b]{\linewidth}\raggedright
Name
\end{minipage} & \begin{minipage}[b]{\linewidth}\raggedright
Capture trigger
\end{minipage} & \begin{minipage}[b]{\linewidth}\raggedright
Challenge window
\end{minipage} & \begin{minipage}[b]{\linewidth}\raggedright
Typical goods
\end{minipage} \\
\midrule\noalign{}
\endhead
\bottomrule\noalign{}
\endlastfoot
D & Deterministic & On-chain verifiable outcome & none & on-chain assets, some digital \\
C & Data commitment & Connector-signed normalised receipt (Section 4.1) whose hash is committed on-chain; the buyer can audit the receipt against the engine record & short (e.g.~24 h) & shipped orders, confirmed appointments, travel \\
A & Attested & Seller-signed attestation without independently checkable record & medium (e.g.~72 h) & on-site services, custom orders, invoices \\
S & Subjective & Buyer acknowledgement or arbiter ruling & until ruling & freelance deliverables, quality-dependent services \\
\end{longtable}

At rungs C and A, capture is \emph{optimistic}: the escrow pays the seller at capture, but the charge remains challengeable for the window, during which the buyer may present evidence (an engine record that does not match the committed hash, a cancellation confirmation) to the arbiter, who may direct a refund from the liquidity order above. At rung S, capture is deferred until acknowledgement or ruling. The rung is part of the \texttt{Offer}, so the buyer agent knows the trust position before signing.

Rung C must be described carefully, because it is the rung most physical-goods and travel purchases live on. Committing a hash of \texttt{(orderId,\ fulfilmentRef)} on-chain does \emph{not} prove that the carrier shipped the parcel, the technician completed the visit, or the airline issued the ticket. It proves that the connector, whose signature is attributable, asserted at a specific time that the engine reported issuance with those references, and it fixes that assertion so it cannot later be altered. Rung C is therefore an \emph{auditable data-commitment} model, not cryptographically verified fulfilment: the buyer, an arbiter or an auditor can retrieve the e-ticket or confirmation from the supplier and check it against the committed receipt. Cryptographic verification of fulfilment would require the supplier or carrier itself to sign, which is rung D and which almost no order-management, scheduling or booking system offers today.

This is nonetheless where A6 is reduced in practice. At rung C an operator that captures without issuance has signed a receipt for a record that does not exist; the mismatch is demonstrable within the challenge window and the refund is sourced from the operator's own pool. The operator's discretion is bounded by its own attributable liability rather than by the buyer's goodwill.

\hypertarget{fees-and-revenue-share}{%
\subsection{Fees and revenue share}\label{fees-and-revenue-share}}

Capture is the single point at which a fee is taken. The fee is a basis-point rate \(f\) on the charge amount, tiered by the seller's trailing volume and capped in absolute terms per charge, and it is split between the operator and, if present, the distributor at share \(s\) of the fee. For a USD 512.40 charge at \(f = 150\) bps and \(s = 60\%\): fee USD 7.69, distributor USD 4.61, operator USD 3.08, seller USD 504.71. Rounding is to the token's smallest unit, with the remainder allocated to the operator so that the three parts sum exactly to the fee.

\textbf{Proposition 2 (self-dealing is unprofitable).} Suppose a distributor can act as, or collude with, a buyer whose spend it refers. If \(s < 1\) and the distributor bears the full charge amount with no subsidy from any party, then each referred charge yields the distributor a net loss of \(f \cdot \text{amount} \cdot (1 - s) > 0\).

\emph{Argument.} The distributor pays \texttt{amount}, receives \texttt{amount\ -\ fee} back as seller (or the seller does, in which case the distributor's loss is larger), and earns \(s \cdot \text{fee}\). Net: \(-\text{fee} + s \cdot \text{fee} < 0\) whenever \(s < 1\). \(\square\)

The condition matters. If a third party subsidises volume (a launch incentive paid per referred charge, for instance), the inequality can flip; such incentives must be sized below \(f(1-s)\) per charge or the guarantee lapses.

\hypertarget{multi-stablecoin-single-currency-charges}{%
\subsection{Multi-stablecoin, single-currency charges}\label{multi-stablecoin-single-currency-charges}}

ASP is currency-agnostic in the sense that USDC is the first entry in a currency registry, and adding another stablecoin is a registry entry plus, where needed, a deposit adapter for the token's authorization method. It is single-currency in the sense that a given charge is denominated, escrowed, captured and refunded in one token. The protocol never converts.

\textbf{Invariant I1.} For every charge, \texttt{currency} is constant across \texttt{deposit}, \texttt{capture} and \texttt{refund}.

The consequence is that there is no DEX, no price oracle, no slippage and no swap-failure state anywhere in the settlement path. If a buyer and seller share no currency, the buyer agent funds one they both accept before quoting; that is a wallet-layer concern, not a settlement one. The cost is that liquidity is fragmented per token and a seller must maintain a vault per accepted currency.

\hypertarget{settlement-modes-and-pause}{%
\subsection{Settlement modes and pause}\label{settlement-modes-and-pause}}

\begin{itemize}
\tightlist
\item
  \textbf{Escrow mode (default).} Funds are held until capture. Safest for the buyer; the seller waits for fulfilment to be paid.
\item
  \textbf{Instant-reserve mode (vetted sellers).} The seller is paid at authorization; refunds are drawn from a rolling reserve funded by a percentage of the seller's captures, with clawback rights. As noted in Section 3, this mode is custodial and requires business verification.
\end{itemize}

A \textbf{pause} role can halt \texttt{deposit} and \texttt{capture} for a token (depeg), a seller (dispute or misbehaving hook) or the whole vault. \texttt{reclaim} is never pausable, so a pause can never strand buyer funds. The pause role is held by a time-locked multisig in the reference deployment (Section 7).

\hypertarget{interface-specification-orchestration-and-conformance}{%
\subsection{Interface specification, orchestration and conformance}\label{interface-specification-orchestration-and-conformance}}

Sections 4.1--4.10 define the protocol; Appendices C--F define the interfaces precisely enough to implement it. The key words MUST, MUST NOT, SHOULD and MAY in the appendices are to be interpreted as in RFC 2119 \citep{rfc2119}. Four interfaces are normative:

\begin{itemize}
\tightlist
\item
  the \textbf{x402 \texttt{asp} scheme} (Appendix C): how an Offer is advertised in a 402 response and how a \texttt{ChargeAuthorization} is returned;
\item
  the \textbf{vault interface} (Appendix D): the on-chain functions, structs, events, errors, access control and revocation semantics every ASP escrow MUST expose;
\item
  the \textbf{operator and connector APIs} (Appendix E): the HTTP surface a buyer agent talks to, the adapter surface a fulfilment-engine connector implements including its timeout rule, the registry and webhook schemas, and the clock and relayer requirements;
\item
  \textbf{conformance levels} (Appendix F): what an implementation MUST support to call itself ASP-Lite, ASP-Core or ASP-Full, demonstrated by an executable suite.
\end{itemize}

Three design rules govern the appendices. Every message that affects settlement is either an on-chain call or a signed off-chain object whose hash is on-chain (Offer, ChargeAuthorization, FulfilmentReceipt); nothing that only lives in an HTTP body can change who gets paid. Every off-chain call is idempotent on \texttt{chargeId}. And every timing field is an integer of UTC unix seconds in chain time, with the connector responsible for converting engine-local time under the \(\delta\) bound.

\textbf{Orchestration.} The on-chain state machine (Section 4.2) is deliberately small. An operator implementation needs a richer off-chain one --- in particular an \texttt{AUTHORIZED\_PENDING\_ENGINE\_HOLD} state that blocks commit until the \texttt{holdBeforeAuth} H1 check has passed, and a \texttt{COMMIT\_UNKNOWN} state entered when the engine call times out, from which the only exit is a \texttt{getHold} query rather than a retry. Because the chain is authoritative for payment state and the engine for fulfilment state, the operator MUST run reconciliation over every non-terminal charge and resolve disagreements toward the authoritative source: engine committed and chain authorized → capture; engine cancelled and chain authorized → void; chain reclaimed and engine hold alive → release the hold; capture transaction status unknown → read the chain before resubmitting. The full off-chain state machine, reconciliation matrix, risk-engine algorithm and relayer design are specified in the companion \emph{ASP-Lite Implementation Specification}, which is the document an engineering team builds from; this paper fixes only what two implementations must agree on to interoperate.

\hypertarget{security-properties}{%
\section{Security Properties}\label{security-properties}}

We state five properties and argue each informally. A mechanised check of P1 and P2 against a TLA+ model of the state machine is planned \citep{tla}.

\textbf{P1 (No stranded funds).} Every deposited amount is eventually either captured to the seller or returned to the buyer, without depending on operator liveness. \emph{Argument.} From \texttt{authorized}, either the operator captures or voids, or \texttt{holdExpiresAt} passes and \texttt{reclaim} becomes callable by anyone (A4). Pause does not block \texttt{reclaim}. From \texttt{captured}, funds are the seller's by construction. No other state holds funds.

\textbf{P2 (No double charge).} A buyer authorization is debited at most once. \emph{Argument.} \texttt{chargeId} is the hash of the authorization struct including a nonce; the escrow rejects a second \texttt{deposit} with the same key, and the token-level authorization (ERC-3009 nonce or Permit2 nonce) is single-use.

\textbf{P3 (Bounded buyer exposure).} A buyer's loss from a malicious operator is bounded by the sum of charges captured without fulfilment within one challenge window, minus what the operator's refund pool covers, and by the smart wallet's spend cap. \emph{Argument.} The operator cannot move escrowed funds anywhere but the seller's registered vault and the fee splitter; capture without fulfilment at rung C is challengeable and refunded from the liquidity order; the wallet cap (Section 7) bounds total authorizations irrespective of operator behaviour.

\textbf{P3' (Late-capture loss assignment).} Protocol-level loss from a capture that fails to land before \(t_h\) is assigned to the operator (Section 4.5, residual risk) and counted in the operator's aggregate liability \(L\). The seller is not exposed to chain timing; it is exposed to the operator's solvency, which Proposition 3 bounds. \emph{Argument.} Commit before \(t_{\text{issue}}\) is the only path to a seller-committed, buyer-reclaimed charge; the operator has undertaken to pay such charges from its pool; and \(C_M\) in \(L\) reserves for them.

\textbf{P4 (Seller payment finality).} Once \texttt{captured}, the seller's net amount cannot be reduced except by a \texttt{refund} within \(W_r\) that the seller's own engine rules authorise or an arbiter directs. \emph{Argument.} The escrow only accepts \texttt{refund} before the window closes and only for amounts up to the captured total; there is no other reducing operation.

\textbf{P5 (Hook isolation).} No seller-supplied code path can move, hold or redirect funds. \emph{Argument.} Off-chain hooks are webhooks with no on-chain authority. On-chain hooks, where enabled, are called by the vault with a read-only view of the charge and can only return an observe/annotate/veto result; the vault, not the hook, executes any transfer, and veto rights are granted per hook by the operator after review.

\hypertarget{integration-with-fulfilment-engines}{%
\section{Integration with Fulfilment Engines}\label{integration-with-fulfilment-engines}}

ASP does not replace a seller's system and does not require changes to it. It sits in front of one, as the agent-facing payment layer, and maps its states onto the engine's own lifecycle through a connector that the operator runs. The claim is deliberately scoped: the seller's \emph{core engine} is unchanged; the seller's \emph{operations} are not (Section 6.4).

\hypertarget{fulfilment-class-registry}{%
\subsection{Fulfilment-class registry}\label{fulfilment-class-registry}}

The protocol defines one lifecycle. Business types differ only in how the four engine operations are named, what the hold time limit is, which verification rung applies, and how refunds are computed. ASP captures those differences in a registry of \emph{fulfilment classes}; an \texttt{Offer} names its class, and the connector for that class implements the mapping. Adding a business type is a registry row and a connector, not a protocol change.

\begin{longtable}[]{@{}
  >{\raggedright\arraybackslash}p{(\columnwidth - 2\tabcolsep) * \real{0.2473}}
  >{\raggedright\arraybackslash}p{(\columnwidth - 2\tabcolsep) * \real{0.7527}}@{}}
\toprule\noalign{}
\begin{minipage}[b]{\linewidth}\raggedright
Registry field
\end{minipage} & \begin{minipage}[b]{\linewidth}\raggedright
Meaning
\end{minipage} \\
\midrule\noalign{}
\endhead
\bottomrule\noalign{}
\endlastfoot
\texttt{classId} & e.g.~\texttt{order.shipped}, \texttt{service.appointment}, \texttt{invoice.b2b}, \texttt{deliverable.accepted}, \texttt{travel.air}, \texttt{travel.hotel} \\
\texttt{holdOp}, \texttt{commitOp}, \texttt{cancelOp}, \texttt{refundQuoteOp} & the engine calls the connector maps ASP transitions onto \\
\texttt{holdLimitSource} & where \(t_e\) is read from (a field, a policy default, an invoice due date) \\
\texttt{minHold} & whether the class can run in escrow mode or must use \texttt{charge} (Section 4.5) \\
\texttt{holdBeforeAuth}, \texttt{policyExpiry} & whether the engine can hold before payment authorization; if not, the conservative \(t_e^{*}\) used at quote time (Section 4.4) \\
\texttt{defaultRung} & verification rung (Section 4.7) \\
\texttt{fulfilmentRefKind} & what the receipt's \texttt{fulfilmentRef} is (tracking number, completion id, ticket number) \\
\texttt{refundRuleSource} & who computes the refundable amount (engine, policy table, arbiter) \\
\texttt{challengeWindow}, \texttt{refundWindow} & class defaults for \(W_c\) and \(W_r\) \\
\end{longtable}

\hypertarget{worked-mappings}{%
\subsection{Worked mappings}\label{worked-mappings}}

The table below gives the mapping for five representative classes. Travel is included because it has the tightest holds and the most complex refund rules; the other columns are what most online service businesses will use.

\begin{longtable}[]{@{}
  >{\raggedright\arraybackslash}p{(\columnwidth - 10\tabcolsep) * \real{0.1087}}
  >{\raggedright\arraybackslash}p{(\columnwidth - 10\tabcolsep) * \real{0.1848}}
  >{\raggedright\arraybackslash}p{(\columnwidth - 10\tabcolsep) * \real{0.1848}}
  >{\raggedright\arraybackslash}p{(\columnwidth - 10\tabcolsep) * \real{0.1576}}
  >{\raggedright\arraybackslash}p{(\columnwidth - 10\tabcolsep) * \real{0.1630}}
  >{\raggedright\arraybackslash}p{(\columnwidth - 10\tabcolsep) * \real{0.2011}}@{}}
\toprule\noalign{}
\begin{minipage}[b]{\linewidth}\raggedright
ASP
\end{minipage} & \begin{minipage}[b]{\linewidth}\raggedright
E-commerce order (OMS)
\end{minipage} & \begin{minipage}[b]{\linewidth}\raggedright
Service appointment
\end{minipage} & \begin{minipage}[b]{\linewidth}\raggedright
B2B invoice
\end{minipage} & \begin{minipage}[b]{\linewidth}\raggedright
Freelance deliverable
\end{minipage} & \begin{minipage}[b]{\linewidth}\raggedright
Air travel (GDS / NDC)
\end{minipage} \\
\midrule\noalign{}
\endhead
\bottomrule\noalign{}
\endlastfoot
engine hold (step 1) & reserve stock & tentative slot hold & invoice issued & task opened, milestone funded & create PNR / \texttt{OrderCreate} \\
\(t_e\) source & stock reservation timeout & slot hold expiry & due date & delivery deadline & ticketing / order time limit \\
commit / \texttt{capture} & shipment confirmed & appointment confirmed or completed & goods/services accepted & deliverable accepted & ticket issued \\
\texttt{void}/\texttt{reclaim} & cancel order, release stock & release slot & invoice cancelled & task cancelled & cancel PNR before ticketing \\
\texttt{refund} & per returns policy & per notice-period schedule & per credit note / dispute & per arbiter or partial acceptance & per fare rules / \texttt{OrderChange} \\
\texttt{fulfilmentRef} & tracking number & completion / visit id & acceptance reference & acceptance hash & e-ticket number \\
default rung & C & C (confirmed) / A (completed) & A & S & C \\
\end{longtable}

In every column the engine hold is created at quote time (step 1 of Section 4.4) so that \(t_e\) is known before the Offer is signed; where a platform cannot hold before payment, the \texttt{holdBeforeAuth} path applies. Two mappings deserve comment. A \emph{service appointment} has two candidate commit points: confirmation (the slot is booked) and completion (the work was done). The class chooses one; confirmation is rung C and captures earlier, completion is rung A and protects the buyer more, and a class may use two charges (deposit at confirmation, balance at completion) to get both. A \emph{B2B invoice} inverts the usual timing: fulfilment often precedes payment, so \texttt{deposit} and \texttt{capture} may be close together and the escrow mainly provides the dispute window before capture.

The connector's hard requirements are: obtain \(t_e\) before quoting (or apply the \texttt{holdBeforeAuth} path), derive \(t_h\) and \(t_{\text{issue}}\) under H1 and place them in the Offer, release provisional holds on decline, and gate commit on \(t_{\text{issue}}\) (H2). H3 is enforced by the vault. H4 is a liveness goal it should also meet.

\hypertarget{extensibility}{%
\subsection{Extensibility}\label{extensibility}}

Sellers often want their own logic at settlement points: issue loyalty points, sync a back office, gate on live inventory. Two tiers are offered. \textbf{Off-chain hooks} are webhooks emitted at each state transition; they run on the seller's infrastructure, cannot affect settlement, and need no review. \textbf{On-chain hooks} are contracts the vault calls at \texttt{capture} and \texttt{refund} for logic that must be atomic with settlement; they are inert until audited and whitelisted, and are constrained by P5. The reference deployment ships off-chain hooks only.

\hypertarget{onboarding-paths}{%
\subsection{Onboarding paths}\label{onboarding-paths}}

\begin{longtable}[]{@{}
  >{\raggedright\arraybackslash}p{(\columnwidth - 4\tabcolsep) * \real{0.3133}}
  >{\raggedright\arraybackslash}p{(\columnwidth - 4\tabcolsep) * \real{0.3855}}
  >{\raggedright\arraybackslash}p{(\columnwidth - 4\tabcolsep) * \real{0.3012}}@{}}
\toprule\noalign{}
\begin{minipage}[b]{\linewidth}\raggedright
Path
\end{minipage} & \begin{minipage}[b]{\linewidth}\raggedright
Change to core booking engine
\end{minipage} & \begin{minipage}[b]{\linewidth}\raggedright
Indicative time to live
\end{minipage} \\
\midrule\noalign{}
\endhead
\bottomrule\noalign{}
\endlastfoot
Platform connector (Shopify, Square, a scheduling SaaS, a GDS/NDC gateway) & none & days \\
Gateway-wrap the seller's own API & none & days to two weeks \\
Direct integration & optional, for control & two to four weeks \\
\end{longtable}

Platform connectors are the route to scale: one connector for a commerce or scheduling platform onboards every seller on it. ``No change to the core engine'' is not ``no work''. Every path requires engine or API access and credentials, a one-time field mapping, cancellation and refund permissions for the connector, a payout vault, business verification, and, for some classes, sector accreditation (travel being the obvious case). The last two are usually the long pole. A fixed sandbox certification (Appendix B) must pass before a seller goes live. Buyer agents require no seller-specific integration. ASP is advertised as an additional scheme in the x402 402 response, so support is implemented once at the x402 client, wallet or facilitator layer (parsing the Offer, constructing the \texttt{ChargeAuthorization}, and tracking the charge's state), after which every ASP seller is reachable without further work.

\hypertarget{instantiation-on-the-xdc-network}{%
\section{Instantiation on the XDC Network}\label{instantiation-on-the-xdc-network}}

The reference deployment targets the XDC Network, an EVM-compatible chain with deterministic finality under XDPoS 2.0 \citep{xdpos2}, which supplies the bounded \(T_f\) that A1 and Proposition 1 rely on. The settlement token is USDC on XDC, which supports ERC-3009, so the default deposit adapter is \texttt{transferWithAuthorization} and the buyer agent never pays gas. Gas for \texttt{deposit}, \texttt{capture}, \texttt{void} and \texttt{refund} is paid by the operator's relayer, which is the same sponsored-gas facilitator infrastructure XDC AI runs for x402 settlement.

Buyer agents hold funds in an ERC-4337 smart account \citep{erc4337} with a session-key policy that enforces a per-charge and rolling spend cap on-chain, so that a compromised agent has a bounded blast radius (P3). Sellers receive proceeds in a counterfactually deployed smart account so that onboarding does not require the seller to hold a key before first capture. Agent and seller identities are registered under ERC-8004 \citep{erc8004}, which supplies the identity and reputation registries the distributor and verification-rung logic reference; we note the empirical caveats about ERC-8004 ecosystem maturity raised by Xiong et al. \citep{erc8004study}.

The pause role is a two-of-three multisig behind a 24-hour timelock, except for the token-level depeg pause, which is immediate and affects \texttt{deposit} only. The vault is intended to be bytecode-compatible with CPP's escrow so that audits of the upstream contract carry over; deviations, if any, will be listed in the deployment notes.

\hypertarget{reference-implementation-scope-asp-lite}{%
\subsection{Reference implementation scope: ASP-Lite}\label{reference-implementation-scope-asp-lite}}

The full profile above is more than a first implementation should attempt. The reference implementation, ASP-Lite, is deliberately narrow so that the safety model can be validated before any of the optional machinery is added:

\begin{itemize}
\tightlist
\item
  USDC on XDC only; one currency-registry entry.
\item
  One seller, one fulfilment class and one connector (the first is expected to be an e-commerce or service-scheduling platform, with a travel NDC connector second), rung C only.
\item
  Escrow mode only; no instant-reserve mode, no rolling reserve beyond what the exposure controls require.
\item
  The seven operations \texttt{quote}, \texttt{authorize}, \texttt{fulfil}, \texttt{capture}, \texttt{void}, \texttt{reclaim}, \texttt{refund}; no partial capture.
\item
  Fee to a single operator address; no distributor split.
\item
  Off-chain hooks only; no on-chain hooks.
\item
  No ERC-8004 registration; identities are operator-issued.
\end{itemize}

Multi-stablecoin support, distributor revenue share, instant-reserve mode, rung A and S arbitration, on-chain hooks and ERC-8004 integration are each added only after ASP-Lite has passed the evaluation in Section 9 on the narrow case.

\hypertarget{related-work}{%
\section{Related Work}\label{related-work}}

\textbf{Industry protocols.} x402 and its schemes \citep{x402, x402schemes}, CPP \citep{cpp}, x402r \citep{x402r}, Cloudflare's deferred scheme \citep{cloudflare402}, AP2 \citep{ap2} and ACP \citep{acp} are discussed in Section 2. Table 5 summarises the positioning.

\begin{longtable}[]{@{}
  >{\raggedright\arraybackslash}p{(\columnwidth - 10\tabcolsep) * \real{0.1985}}
  >{\raggedright\arraybackslash}p{(\columnwidth - 10\tabcolsep) * \real{0.1397}}
  >{\raggedright\arraybackslash}p{(\columnwidth - 10\tabcolsep) * \real{0.1618}}
  >{\raggedright\arraybackslash}p{(\columnwidth - 10\tabcolsep) * \real{0.1176}}
  >{\raggedright\arraybackslash}p{(\columnwidth - 10\tabcolsep) * \real{0.1471}}
  >{\raggedright\arraybackslash}p{(\columnwidth - 10\tabcolsep) * \real{0.2353}}@{}}
\toprule\noalign{}
\begin{minipage}[b]{\linewidth}\raggedright
Layer
\end{minipage} & \begin{minipage}[b]{\linewidth}\raggedright
x402 (\texttt{exact})
\end{minipage} & \begin{minipage}[b]{\linewidth}\raggedright
CPP / x402r
\end{minipage} & \begin{minipage}[b]{\linewidth}\raggedright
AP2
\end{minipage} & \begin{minipage}[b]{\linewidth}\raggedright
ACP
\end{minipage} & \begin{minipage}[b]{\linewidth}\raggedright
ASP
\end{minipage} \\
\midrule\noalign{}
\endhead
\bottomrule\noalign{}
\endlastfoot
Standardises & metered access & escrow operations & consent chain & checkout messages & fulfilment-coupled settlement \\
Rail & stablecoin & stablecoin & any & cards / processor & stablecoin \\
Hold / capture & no & yes & n/a & via processor & yes (CPP-compatible) \\
Hold tied to inventory & n/a & no & n/a & processor-specific & yes (H1--H3, issuance deadline) \\
Who asserts fulfilment & n/a & operator, unspecified & n/a & merchant & per-rung, specified \\
Refund amount source & n/a & operator-specified & n/a & merchant & engine's fare / rate rules \\
Revenue share & n/a & fee range only & n/a & none & at capture, with Prop. 2 \\
\end{longtable}

\textbf{Academic literature.} Zhang et al.~systematise blockchain agent-to-agent payments around discovery, authorisation, execution and accounting, and note the absence of reversibility in shipped designs \citep{sokA2A}. Mao et al.~survey the security of LLM agents in agentic commerce and catalogue attacks on the payment step in particular \citep{sokAgenticCommerce}. A402 binds a cryptocurrency payment to verifiable service execution through an atomic service channel, which is the rung-D case of our ladder \citep{a402}. The systematic analysis of x402 free-riding shows how a seller can be induced to serve without payment or a buyer to pay without service under the \texttt{exact} scheme, which is the failure mode escrow removes \citep{freeriding}. See and Tan integrate programmable compliance into a stablecoin rail and use an escrow tranche for pending compliance conditions; their compliance hooks are complementary to ASP's settlement hooks \citep{compliance}. Lan et al.'s analysis of AP2 replay is the reason ASP binds the authorization struct to \texttt{offerId}, \texttt{holdExpiresAt} and a nonce \citep{ap2replay}. Vaziry et al.~combine A2A with x402 micropayments and ledger-anchored identities \citep{vaziry}; AESP proposes privacy-preserving settlement for agents \citep{aesp}; Alqithami surveys execution models and trust boundaries for on-chain agents \citep{alqithami}. The earliest agent-mediated payment protocols predate blockchains by two decades \citep{pang2002} and already identified the need to separate authorization from settlement.

To our knowledge no prior work specifies the coupling between an on-chain hold and an off-chain fulfilment engine's hold, or states the fulfilment-verification trust position per class of good or service.

\hypertarget{limitations-and-open-questions}{%
\section{Limitations and Open Questions}\label{limitations-and-open-questions}}

\begin{itemize}
\tightlist
\item
  \textbf{No deployed measurement.} Gas per operation, latency to finality on XDC, and behaviour under expiry races are unmeasured. The evaluation plan below is the most important follow-up, and this paper should be read as the specification that plan tests.
\item
  \textbf{Operator trust at rungs A and S.} The challenge window bounds but does not eliminate operator discretion. Multi-operator or arbiter-set designs are future work.
\item
  \textbf{Partial capture.} Multi-leg itineraries and split deliveries need partial capture, which CPP supports and this profile does not yet specify.
\item
  \textbf{Privacy.} Charge amounts, seller vaults and timing are public on-chain. Mitigations (amount blinding, batched capture) are out of scope here.
\item
  \textbf{Regulatory posture.} Sector accreditation where it applies (IATA for air), consumer-protection rules on refunds and cancellations that vary by class and jurisdiction, stablecoin regimes (MiCA in the EU, VARA in the UAE) and the classification of the operator's refund pool are unresolved and jurisdiction-dependent.
\item
  \textbf{Pricing model.} Whether ASP is priced as a thin rail or as a distribution channel sets the fee ceiling and is a commercial decision outside the protocol.
\item
  \textbf{Name.} ``Agentic Settlement Protocol'' was chosen to avoid collision with ``Agentic Commerce Protocol''; it remains a working name.
\end{itemize}

\hypertarget{evaluation-plan}{%
\subsection{Evaluation plan}\label{evaluation-plan}}

ASP-Lite will be exercised on XDC testnet, then mainnet, against a simulated fulfilment engine that implements the generic hold / commit / cancel / refund-quote lifecycle with a configurable hold time limit (so that short-hold scheduling classes and long-hold travel classes are both covered), and against one real connector. Each run executes a booking lifecycle end to end; fault injection is applied per run from the following catalogue: engine timeout before issuance; engine commit after \(t_{\text{issue}}\) (must be refused by H2); operator crash between attestation and capture; duplicate engine callbacks; duplicate \texttt{capture} and duplicate \texttt{refund} submissions; RPC outage of configurable duration during the capture window; capture broadcast within \(T_s\) of \(t_{\text{issue}}\); \texttt{reclaim} submitted at \(t_h \pm\) one block; refund request when the seller vault is empty; refund request at the refund-window boundary; token pause during an open hold.

Metrics recorded per run: time from attestation to capture inclusion and to finality; time from refund request to refund finality; captures that reverted and why; late-capture losses (issued, then reclaimed); reclaim frequency and cause; H2 refusals; H3 reverts; \texttt{holdBeforeAuth} void-and-requote events; refund source used (vault, reserve, delayed, pool); gas per operation by role; and, across runs, the empirical distribution of \(T_s + T_i\) against the configured margin. The target is on the order of thousands of lifecycles, with a fault injected in a stated fraction of them, so that Proposition 1 and the exposure controls are tested against measured rather than assumed delays. The Appendix B cases are packaged as an executable conformance suite (\texttt{asp-conformance}) that any vault or connector implementation can run and that reports the number of cases passed at the claimed level; passing it is the onboarding gate for sellers and connectors. A second, independently written ASP-Lite implementation exercising the same suite against the first's vault will test the interoperability claim of Appendix F. Results will be reported in a revised version of this paper.

\hypertarget{conclusion}{%
\section{Conclusion}\label{conclusion}}

Metered agent payments are solved; agent \emph{purchases} are not, because a purchase needs a hold that lapses with the inventory, a fulfilment assertion someone is accountable for, and a refund computed by whoever owns the fare rules. The escrow primitive that makes this possible on-chain already exists. ASP specifies how to drive it: an issuance deadline that keeps the two holds apart by a margin the operator is accountable for, a verification ladder that says who is trusted, what their assertion proves and for how long, a refund model that defers to the engine and bounds the operator's exposure, and revenue share that cannot be farmed. Travel forces every one of these to exist, which is why it is the running example; but the protocol is defined against an abstract fulfilment engine, and physical orders, service appointments, invoices, freelance deliverables and subscriptions are each a row in the fulfilment-class registry over the same machine.

\hypertarget{appendix-a-authorization-struct}{%
\appendix
\section{Authorization struct}\label{appendix-a-authorization-struct}}

\begin{verbatim}
struct Offer {               // signed by seller, delegate or operator (Section 4.3)
  bytes32 sellerId;
  bytes32 fulfilmentClass;
  bytes32 engineRef;
  address token;
  uint256 amount;
  uint64  engineExpiry;      // t_e
  uint64  issueDeadline;     // t_issue
  uint64  holdExpiresAt;     // t_h
  uint8   verificationRung;
  uint32  challengeWindow;   // seconds
  uint32  refundWindow;      // seconds
  bytes32 refundPolicyRef;
  uint64  expiresAt;
  address signer;
  bytes32 signerAuthority;   // direct | delegation id | mandate id
  uint32  delegationVersion; // seller's delegationVersion at signing (0 if direct)
}
// offerId = keccak256(EIP-712 encoding of Offer); binds every term above.

struct ChargeAuthorization { // signed by the buyer agent
  bytes32 offerId;
  address buyer;
  address token;
  uint256 amount;
  uint64  issueDeadline;     // must equal Offer.issueDeadline
  uint64  holdExpiresAt;     // must equal Offer.holdExpiresAt
  bytes32 nonce;
}
// EIP-712 domain: name "ASP", version "1", chainId, verifyingContract = vault
// chargeId = keccak256(abi.encode(TYPEHASH, ...fields))
// deposit() receives the Offer terms alongside, recomputes offerId, and reverts
// on mismatch; it stores issueDeadline and holdExpiresAt for capture()/reclaim().
\end{verbatim}

\hypertarget{appendix-b-sandbox-certification-cases}{%
\section{Sandbox certification cases}\label{appendix-b-sandbox-certification-cases}}

\begin{enumerate}
\def\labelenumi{\arabic{enumi}.}
\tightlist
\item
  Happy path: quote, deposit, fulfil, capture; seller receives \texttt{amount\ -\ fee}; splitter receives fee in correct shares.
\item
  Void: deposit, engine fails to issue, void; buyer receives full amount.
\item
  Expiry: deposit, operator silent, \texttt{reclaim} by third party after \texttt{holdExpiresAt}; buyer receives full amount; capture afterwards reverts.
\item
  Refund: capture, engine returns partial refundable amount, refund for exactly that amount; refund beyond window reverts; cumulative refund cannot exceed capture.
\item
  Idempotency: duplicate deposit, duplicate capture, duplicate refund are no-ops with identical results.
\item
  H2 gate: issuance request after \(t_{\text{issue}}\) is refused by the connector even though the engine reservation is alive.
\item
  H3 gate: a \texttt{capture} whose attestation has \texttt{issuedAt} \(> t_{\text{issue}}\) reverts in the vault; the operator's pre-check also refuses to submit it.
\item
  H4 (engine first): engine hold lapses, connector voids promptly.
\item
  H4 (escrow first): escrow hold lapses, connector cancels engine hold; no issuance possible.
\item
  Exposure cap: a capture that would push a seller's uncovered exposure above \(E_{\max}\) is refused before deposit.
\item
  Pause: token pause blocks deposit and capture, does not block reclaim.
\item
  Offer binding: a deposit whose authorization timing fields differ from the Offer's reverts; an Offer signed by a revoked delegate is rejected.
\item
  Decline: buyer never authorizes; provisional engine hold is released at \texttt{expiresAt}.
\item
  \texttt{holdBeforeAuth\ =\ false} shortfall: engine returns \(t_e < t_h + \delta\) after deposit; connector voids immediately and re-quotes.
\item
  Partial refund: two refunds summing to less than capture leave the charge \texttt{captured}; a third bringing the cumulative total to the captured amount moves it to \texttt{refunded}; any further refund reverts.
\item
  Commit timeout: the engine accepts the commit but the response is lost; the operator observes \texttt{COMMIT\_UNKNOWN}, calls \texttt{getHold}, finds \texttt{committed}, and captures exactly once with no second engine call.
\item
  Exposure reservation: with headroom for one Offer, two concurrent quote requests result in exactly one Offer issued and one \texttt{EXPOSURE\_CAP} refusal; the reservation is released when the issued Offer expires unauthorized.
\end{enumerate}

\hypertarget{appendix-c-the-x402-asp-scheme}{%
\section{\texorpdfstring{Appendix C: The x402 \texttt{asp} scheme}{Appendix C: The x402 asp scheme}}\label{appendix-c-the-x402-asp-scheme}}

\textbf{C.1 Advertising.} A seller endpoint that supports ASP MUST include, in the \texttt{accepts} array of its x402 402 response, an entry with \texttt{scheme:\ "asp"}. The entry carries the signed Offer and enough context for the client to construct a \texttt{ChargeAuthorization}. Networks are identified by CAIP-2 \citep{caip2}.

\begin{Shaded}
\begin{Highlighting}[]
\FunctionTok{\{}
  \DataTypeTok{"x402Version"}\FunctionTok{:} \DecValTok{2}\FunctionTok{,}
  \DataTypeTok{"accepts"}\FunctionTok{:} \OtherTok{[}\FunctionTok{\{}
    \DataTypeTok{"scheme"}\FunctionTok{:} \StringTok{"asp"}\FunctionTok{,}
    \DataTypeTok{"network"}\FunctionTok{:} \StringTok{"eip155:50"}\FunctionTok{,}
    \DataTypeTok{"asset"}\FunctionTok{:} \StringTok{"0x\textless{}usdc{-}on{-}xdc\textgreater{}"}\FunctionTok{,}
    \DataTypeTok{"amount"}\FunctionTok{:} \StringTok{"512400000"}\FunctionTok{,}
    \DataTypeTok{"payTo"}\FunctionTok{:} \StringTok{"0x\textless{}vault\textgreater{}"}\FunctionTok{,}
    \DataTypeTok{"maxTimeoutSeconds"}\FunctionTok{:} \DecValTok{900}\FunctionTok{,}
    \DataTypeTok{"extra"}\FunctionTok{:} \FunctionTok{\{}
      \DataTypeTok{"aspVersion"}\FunctionTok{:} \StringTok{"1"}\FunctionTok{,}
      \DataTypeTok{"offer"}\FunctionTok{:} \FunctionTok{\{} \DataTypeTok{"...Offer fields, Appendix A..."} \FunctionTok{\},}
      \DataTypeTok{"offerId"}\FunctionTok{:} \StringTok{"0x\textless{}keccak256 of EIP{-}712 Offer\textgreater{}"}\FunctionTok{,}
      \DataTypeTok{"offerSignature"}\FunctionTok{:} \StringTok{"0x\textless{}65 bytes\textgreater{}"}\FunctionTok{,}
      \DataTypeTok{"depositMethod"}\FunctionTok{:} \StringTok{"eip3009"}\FunctionTok{,}
      \DataTypeTok{"eip712Domain"}\FunctionTok{:} \FunctionTok{\{} \DataTypeTok{"name"}\FunctionTok{:} \StringTok{"ASP"}\FunctionTok{,} \DataTypeTok{"version"}\FunctionTok{:} \StringTok{"1"}\FunctionTok{,}
                        \DataTypeTok{"chainId"}\FunctionTok{:} \DecValTok{50}\FunctionTok{,} \DataTypeTok{"verifyingContract"}\FunctionTok{:} \StringTok{"0x\textless{}vault\textgreater{}"} \FunctionTok{\},}
      \DataTypeTok{"operatorApi"}\FunctionTok{:} \StringTok{"https://\textless{}operator\textgreater{}/asp/v1"}\FunctionTok{,}
      \DataTypeTok{"statusUrl"}\FunctionTok{:} \StringTok{"https://\textless{}operator\textgreater{}/asp/v1/charges/\{chargeId\}"}
    \FunctionTok{\}}
  \FunctionTok{\}}\OtherTok{]}
\FunctionTok{\}}
\end{Highlighting}
\end{Shaded}

Rules: \texttt{amount} MUST equal \texttt{offer.amount}; \texttt{payTo} MUST be the vault; \texttt{maxTimeoutSeconds} MUST be at most \texttt{offer.expiresAt\ -\ now}. A client MUST recompute \texttt{offerId} from \texttt{offer} and MUST reject the entry if it differs from the advertised value or if \texttt{offerSignature} does not recover to \texttt{offer.signer}. A client SHOULD verify \texttt{signerAuthority} against the operator's seller registry (Appendix E.3) before signing.

\textbf{C.2 Payment payload.} The client retries with the standard x402 payment header whose payload is:

\begin{Shaded}
\begin{Highlighting}[]
\FunctionTok{\{}
  \DataTypeTok{"scheme"}\FunctionTok{:} \StringTok{"asp"}\FunctionTok{,}
  \DataTypeTok{"network"}\FunctionTok{:} \StringTok{"eip155:50"}\FunctionTok{,}
  \DataTypeTok{"payload"}\FunctionTok{:} \FunctionTok{\{}
    \DataTypeTok{"authorization"}\FunctionTok{:} \FunctionTok{\{} \DataTypeTok{"...ChargeAuthorization fields, Appendix A..."} \FunctionTok{\},}
    \DataTypeTok{"authorizationSignature"}\FunctionTok{:} \StringTok{"0x\textless{}65 bytes\textgreater{}"}\FunctionTok{,}
    \DataTypeTok{"depositMethod"}\FunctionTok{:} \StringTok{"eip3009"}\FunctionTok{,}
    \DataTypeTok{"depositProof"}\FunctionTok{:} \StringTok{"0x\textless{}ERC{-}3009 / EIP{-}2612 / Permit2 signature\textgreater{}"}
  \FunctionTok{\}}
\FunctionTok{\}}
\end{Highlighting}
\end{Shaded}

\texttt{authorization.offerId} MUST equal the advertised \texttt{offerId}, and \texttt{issueDeadline} and \texttt{holdExpiresAt} MUST equal the Offer's. The facilitator (the operator's relayer) verifies the payload, calls \texttt{deposit}, and responds with the x402 settlement header containing \texttt{chargeId}, the deposit transaction hash and \texttt{statusUrl}. Unlike \texttt{exact}, a successful settlement response means \emph{funds are held}, not paid; the resource is delivered when the charge reaches \texttt{captured}, and the client MUST poll \texttt{statusUrl} or subscribe to webhooks to learn that.

\textbf{C.3 Failure codes.} In addition to x402's own errors, the facilitator returns one of: \texttt{OFFER\_EXPIRED}, \texttt{OFFER\_ID\_MISMATCH}, \texttt{OFFER\_SIGNER\_UNAUTHORISED}, \texttt{TIMING\_MISMATCH} (authorization timing fields differ from Offer), \texttt{H1\_VIOLATION} (only on the \texttt{holdBeforeAuth} path, after void), \texttt{EXPOSURE\_CAP} (seller or operator ceiling breached), \texttt{TOKEN\_PAUSED}.

\hypertarget{appendix-d-vault-interface}{%
\section{Vault interface}\label{appendix-d-vault-interface}}

Every ASP escrow MUST implement the following interface. Amounts are in the token's smallest unit; times are unix seconds.

\begin{Shaded}
\begin{Highlighting}[]
\NormalTok{interface IASPVault \{}
\NormalTok{    enum Status \{ None, Authorized, Captured, Refunded, Reclaimed \}}

\NormalTok{    struct Offer \{ /* Appendix A */ \}}
\NormalTok{    struct ChargeAuthorization \{ /* Appendix A */ \}}
\NormalTok{    struct Attestation \{}
\NormalTok{        bytes32 chargeId; bytes32 receiptHash; uint64 issuedAt;}
\NormalTok{        address attestor; bytes attestorSig;}
\NormalTok{    \}}
\NormalTok{    struct Charge \{}
\NormalTok{        Status status; address buyer; address token; uint256 amount;}
\NormalTok{        uint256 capturedAmount; uint256 refundedAmount;}
\NormalTok{        uint64 issueDeadline; uint64 holdExpiresAt; uint64 refundWindowEnd;}
\NormalTok{        bytes32 offerId; bytes32 receiptHash;}
\NormalTok{    \}}

\NormalTok{    struct Delegation \{            // seller {-}\textgreater{} connector signing key}
\NormalTok{        bytes32 sellerId; address signer; uint32 version;}
\NormalTok{        uint64 validUntil; uint256 maxAmount; bytes32[] classes;}
\NormalTok{    \}}
\NormalTok{    struct AttestorGrant \{         // seller/class {-}\textgreater{} connector attestor key}
\NormalTok{        bytes32 sellerId; bytes32 classId; address attestor;}
\NormalTok{        uint64 validFrom; uint64 validUntil;}
\NormalTok{    \}}

\NormalTok{    // {-}{-}{-} registry (operator role; seller may revoke) {-}{-}{-}}
\NormalTok{    function setDelegation(Delegation calldata d, bytes calldata sellerSig) external;}
\NormalTok{    function revokeDelegations(bytes32 sellerId) external;   // bumps delegationVersion}
\NormalTok{    function delegationVersion(bytes32 sellerId) external view returns (uint32);}
\NormalTok{    function setAttestor(AttestorGrant calldata g) external;}
\NormalTok{    function revokeAttestor(bytes32 sellerId, bytes32 classId, address attestor) external;}

\NormalTok{    // {-}{-}{-} lifecycle {-}{-}{-}}
\NormalTok{    function deposit(Offer calldata offer, bytes calldata offerSig,}
\NormalTok{                     ChargeAuthorization calldata auth, bytes calldata authSig,}
\NormalTok{                     bytes calldata depositProof) external returns (bytes32 chargeId);}
\NormalTok{    function capture(bytes32 chargeId, Attestation calldata att) external;}
\NormalTok{    function void(bytes32 chargeId) external;}
\NormalTok{    function reclaim(bytes32 chargeId) external;   // anyone, after holdExpiresAt}
\NormalTok{    function refund(bytes32 chargeId, uint256 amount, uint8 source) external;}

\NormalTok{    // {-}{-}{-} views {-}{-}{-}}
\NormalTok{    function charges(bytes32 chargeId) external view returns (Charge memory);}
\NormalTok{    function offerId(Offer calldata offer) external view returns (bytes32);}
\NormalTok{    function chargeId(ChargeAuthorization calldata auth) external view returns (bytes32);}

\NormalTok{    // {-}{-}{-} events {-}{-}{-}}
\NormalTok{    event Authorized(bytes32 indexed chargeId, bytes32 indexed offerId,}
\NormalTok{                     address indexed buyer, address token, uint256 amount,}
\NormalTok{                     uint64 issueDeadline, uint64 holdExpiresAt);}
\NormalTok{    event DelegationSet(bytes32 indexed sellerId, address signer, uint32 version);}
\NormalTok{    event DelegationsRevoked(bytes32 indexed sellerId, uint32 newVersion);}
\NormalTok{    event AttestorSet(bytes32 indexed sellerId, bytes32 indexed classId, address attestor,}
\NormalTok{                      uint64 validFrom, uint64 validUntil);}
\NormalTok{    event Captured(bytes32 indexed chargeId, bytes32 receiptHash,}
\NormalTok{                   uint256 toSeller, uint256 fee, uint256 reserve);}
\NormalTok{    event Voided(bytes32 indexed chargeId);}
\NormalTok{    event Reclaimed(bytes32 indexed chargeId, address caller);}
\NormalTok{    event Refunded(bytes32 indexed chargeId, uint256 amount,}
\NormalTok{                   uint256 cumulative, uint8 source);}
\NormalTok{    event Paused(bytes32 scope, bool paused);     // token | seller | vault}

\NormalTok{    // {-}{-}{-} errors {-}{-}{-}}
\NormalTok{    error OfferExpired(); error OfferIdMismatch(); error InvalidOfferSignature();}
\NormalTok{    error InvalidDelegate(uint32 offerVersion, uint32 currentVersion);}
\NormalTok{    error TimingMismatch(); error AuthorizationUsed(); error WrongStatus(Status have);}
\NormalTok{    error IssueDeadlinePassed(uint64 issuedAt, uint64 deadline);}
\NormalTok{    error InvalidAttestor(address attestor, uint64 issuedAt);}
\NormalTok{    error HoldExpired(); error HoldNotExpired();}
\NormalTok{    error RefundWindowClosed(); error RefundExceeded(uint256 cumulative, uint256 captured);}
\NormalTok{    error ExposureCap(); error PausedScope(bytes32 scope); error NotOperator();}
\NormalTok{\}}
\end{Highlighting}
\end{Shaded}

Normative behaviour:

\begin{itemize}
\tightlist
\item
  \texttt{deposit} MUST recompute \texttt{offerId} from \texttt{offer}, verify \texttt{offerSig} recovers to \texttt{offer.signer}, verify that \texttt{signer} is the seller's direct key or holds a Delegation whose \texttt{version\ ==\ delegationVersion(sellerId)}, whose \texttt{validUntil} has not passed, whose \texttt{maxAmount\ \textgreater{}=\ offer.amount} and whose \texttt{classes} include \texttt{offer.fulfilmentClass} (else \texttt{InvalidDelegate}), verify \texttt{authSig}, and revert with \texttt{TimingMismatch} unless \texttt{auth.issueDeadline\ ==\ offer.issueDeadline} and \texttt{auth.holdExpiresAt\ ==\ offer.holdExpiresAt}. It MUST revert with \texttt{HoldExpired} if \texttt{block.timestamp\ \textgreater{}=\ offer.holdExpiresAt\ -\ M\_min} (a vault constant, default 300 s). It MUST be a no-op returning the existing \texttt{chargeId} on a duplicate, and MUST consume the token-level authorization nonce so a second \texttt{deposit} with a different \texttt{chargeId} but the same buyer nonce reverts with \texttt{AuthorizationUsed}.
\item
  \textbf{Revocation semantics.} \texttt{revokeDelegations(sellerId)} increments the seller's \texttt{delegationVersion}. An Offer carries the version under which it was signed (Appendix A, \texttt{delegationVersion}); an Offer whose version is lower than the current one is invalid at \texttt{deposit} \emph{immediately}, even if \texttt{expiresAt} has not passed. Charges already \texttt{Authorized} under the old version are unaffected. This choice favours the seller (a compromised key is neutralised at once) at the cost of buyers who hold not-yet-deposited Offers, which are short-lived by construction.
\item
  \texttt{capture} MUST revert with \texttt{IssueDeadlinePassed} if \texttt{att.issuedAt\ \textgreater{}\ issueDeadline}, MUST verify \texttt{att.attestorSig} recovers to an attestor with an AttestorGrant for \texttt{(sellerId,\ fulfilmentClass)} whose \texttt{{[}validFrom,\ validUntil{]}} contains \texttt{att.issuedAt} (else \texttt{InvalidAttestor}; an attestation signed before a key's revocation but submitted after it remains valid, since validity is judged at \texttt{issuedAt}), and MUST pay \texttt{amount\ -\ fee\ -\ reserve} to the seller vault, \texttt{fee} to the splitter and \texttt{reserve} to the reserve vault in the same transaction. It MUST set \texttt{refundWindowEnd\ =\ block.timestamp\ +\ offer.refundWindow}.
\item
  \texttt{reclaim} MUST NOT be restricted by caller or by any pause, and MUST succeed whenever \texttt{status\ ==\ Authorized\ \&\&\ block.timestamp\ \textgreater{}=\ holdExpiresAt}.
\item
  \texttt{refund} MUST revert if \texttt{block.timestamp\ \textgreater{}\ refundWindowEnd} or \texttt{refundedAmount\ +\ amount\ \textgreater{}\ capturedAmount} (\texttt{RefundExceeded}); it MUST transition to \texttt{Refunded} exactly when the cumulative equals \texttt{capturedAmount}. \textbf{Atomic sourcing.} The vault MUST keep, per seller and token, three internal balances --- \texttt{available}, \texttt{reserve} and \texttt{delayed} (with release schedule) --- plus one operator \texttt{pool} balance, and a single \texttt{refund} call MUST draw the amount from them in that order within one transaction. The \texttt{source} argument is advisory and the \texttt{Refunded} event reports the tier actually reached; if the pool is also insufficient the call reverts and no partial refund is made. Sequential refunds from separate wallets are non-conformant because they create partial-failure states.
\item
  Only the operator role MAY call \texttt{deposit}, \texttt{capture}, \texttt{void} and \texttt{refund}. The pause role MAY pause \texttt{deposit} and \texttt{capture} per scope; it MUST NOT be able to pause \texttt{reclaim}.
\item
  Fee, reserve and settlement-delay parameters MUST be read from the seller's registry entry at capture time, not supplied by the caller.
\item
  \texttt{capture} MUST decrement the seller's \emph{committed-not-captured} exposure and increment \emph{refundable} exposure in the same transaction as the transfer (Section 4.6, reservation model), so that the on-chain view of exposure is never stale relative to funds.
\end{itemize}

\hypertarget{appendix-e-off-chain-interfaces}{%
\section{Off-chain interfaces}\label{appendix-e-off-chain-interfaces}}

\textbf{E.1 Operator API} (base path \texttt{/asp/v1}; JSON; all responses carry \texttt{aspVersion}). Requests that create or change state MUST carry an \texttt{Idempotency-Key} header equal to \texttt{chargeId} where one exists.

\begin{longtable}[]{@{}
  >{\raggedright\arraybackslash}p{(\columnwidth - 6\tabcolsep) * \real{0.0674}}
  >{\raggedright\arraybackslash}p{(\columnwidth - 6\tabcolsep) * \real{0.3146}}
  >{\raggedright\arraybackslash}p{(\columnwidth - 6\tabcolsep) * \real{0.1348}}
  >{\raggedright\arraybackslash}p{(\columnwidth - 6\tabcolsep) * \real{0.4831}}@{}}
\toprule\noalign{}
\begin{minipage}[b]{\linewidth}\raggedright
Method
\end{minipage} & \begin{minipage}[b]{\linewidth}\raggedright
Path
\end{minipage} & \begin{minipage}[b]{\linewidth}\raggedright
Caller
\end{minipage} & \begin{minipage}[b]{\linewidth}\raggedright
Purpose
\end{minipage} \\
\midrule\noalign{}
\endhead
\bottomrule\noalign{}
\endlastfoot
\texttt{POST} & \texttt{/quotes} & buyer agent & Body: \texttt{\{\ sellerId,\ fulfilmentClass,\ items,\ buyer\ \}}. Creates the provisional engine hold, derives deadlines, returns the signed Offer as in Appendix C.1. \\
\texttt{POST} & \texttt{/charges} & buyer agent / facilitator & Body: the C.2 payload. Relays \texttt{deposit}; returns \texttt{\{\ chargeId,\ txHash,\ status\ \}}. \\
\texttt{GET} & \texttt{/charges/\{chargeId\}} & any & Returns the \texttt{Charge} view plus off-chain fields: \texttt{engineRef}, \texttt{receipt} (after capture), \texttt{refunds{[}{]}}, \texttt{challenge} (if any). \\
\texttt{POST} & \texttt{/charges/\{chargeId\}/cancel} & buyer agent & Pre-capture: requests \texttt{void}. Post-capture: submits cancellation to the engine, returns \texttt{\{\ refundableAmount,\ refundPolicyRef,\ expiresAt\ \}} as a quote. \\
\texttt{POST} & \texttt{/charges/\{chargeId\}/refund} & buyer agent & Accepts a refund quote; operator executes \texttt{refund} for exactly \texttt{refundableAmount}. \\
\texttt{GET} & \texttt{/charges/\{chargeId\}/receipt} & buyer, arbiter, auditor & Returns the normalised \texttt{FulfilmentReceipt}, the connector signature, \texttt{receiptHash}, and engine references. MUST be retained for at least \(W_c + W_r\) and SHOULD be retained for the operator's audit period. \\
\texttt{GET} & \texttt{/charges/\{chargeId\}/refunds} & buyer, arbiter, auditor & Returns each refund's engine quote, refundable amount, reason, \texttt{refundPolicyRef}, connector response and timestamp: the off-chain evidence for an on-chain \texttt{refund}. \\
\texttt{POST} & \texttt{/charges/\{chargeId\}/challenge} & buyer agent & Within \(W_c\): body \texttt{\{\ claim,\ evidence{[}{]}\ \}}. Opens a case with the arbiter (E.5). \\
\texttt{GET} & \texttt{/registry/classes} & any & Fulfilment-class registry (E.4). \\
\texttt{GET} & \texttt{/registry/sellers/\{sellerId\}} & any & Public part of the seller registry (E.3): delegations, tier, accepted tokens, vault. \\
\texttt{POST} & \texttt{/webhooks} & seller / buyer agent & Subscribe to E.6 events for a \texttt{sellerId} or \texttt{buyer}. \\
\end{longtable}

Errors use RFC 7807 problem details with \texttt{type} set to one of the C.3 codes or \texttt{ENGINE\_UNAVAILABLE}, \texttt{CHALLENGE\_WINDOW\_CLOSED}, \texttt{REFUND\_QUOTE\_EXPIRED}. A machine-readable OpenAPI 3.1 description \citep{openapi} of this surface accompanies the reference implementation.

\textbf{E.2 Connector interface.} A fulfilment-engine connector is a component the operator loads per fulfilment class. It MUST implement:

\begin{Shaded}
\begin{Highlighting}[]
\KeywordTok{interface}\NormalTok{ Ctx \{ sellerId}\OperatorTok{:} \DataTypeTok{string}\OperatorTok{;}\NormalTok{ offerId}\OperatorTok{?:} \DataTypeTok{string}\OperatorTok{;}\NormalTok{ chargeId}\OperatorTok{?:} \DataTypeTok{string}\OperatorTok{;}
\NormalTok{                engineRef}\OperatorTok{?:} \DataTypeTok{string}\OperatorTok{;}\NormalTok{ idempotencyKey}\OperatorTok{:} \DataTypeTok{string}\OperatorTok{;}\NormalTok{ \}}

\KeywordTok{interface}\NormalTok{ FulfilmentConnector \{}
\NormalTok{  classId}\OperatorTok{:} \DataTypeTok{string}\OperatorTok{;}
  \FunctionTok{createHold}\NormalTok{(ctx}\OperatorTok{:}\NormalTok{ Ctx}\OperatorTok{,}\NormalTok{ req}\OperatorTok{:}\NormalTok{ HoldRequest)}\OperatorTok{:} \BuiltInTok{Promise}\OperatorTok{\textless{}}\NormalTok{\{ engineRef}\OperatorTok{:} \DataTypeTok{string}\OperatorTok{;}
\NormalTok{      engineExpiry}\OperatorTok{:} \DataTypeTok{number} \OperatorTok{|} \DataTypeTok{null}\OperatorTok{;}\NormalTok{ price}\OperatorTok{:} \DataTypeTok{bigint}\OperatorTok{;}\NormalTok{ currency}\OperatorTok{:} \DataTypeTok{string}\NormalTok{ \}}\OperatorTok{\textgreater{};}
  \FunctionTok{getHold}\NormalTok{(ctx}\OperatorTok{:}\NormalTok{ Ctx)}\OperatorTok{:} \BuiltInTok{Promise}\OperatorTok{\textless{}}\NormalTok{\{ state}\OperatorTok{:} \StringTok{"held"} \OperatorTok{|} \StringTok{"committed"} \OperatorTok{|} \StringTok{"cancelled"} \OperatorTok{|} \StringTok{"expired"} \OperatorTok{|} \StringTok{"unknown"}\OperatorTok{;}
\NormalTok{      engineExpiry}\OperatorTok{:} \DataTypeTok{number} \OperatorTok{|} \DataTypeTok{null}\OperatorTok{;}\NormalTok{ record}\OperatorTok{?:}\NormalTok{ EngineRecord \}}\OperatorTok{\textgreater{};}   \CommentTok{// reconciliation}
  \FunctionTok{releaseHold}\NormalTok{(ctx}\OperatorTok{:}\NormalTok{ Ctx)}\OperatorTok{:} \BuiltInTok{Promise}\OperatorTok{\textless{}}\DataTypeTok{void}\OperatorTok{\textgreater{};}                        \CommentTok{// decline / lapse / H4}
  \FunctionTok{commit}\NormalTok{(ctx}\OperatorTok{:}\NormalTok{ Ctx)}\OperatorTok{:} \BuiltInTok{Promise}\OperatorTok{\textless{}}\NormalTok{FulfilmentReceipt}\OperatorTok{\textgreater{};}                 \CommentTok{// before t\_issue}
  \FunctionTok{cancel}\NormalTok{(ctx}\OperatorTok{:}\NormalTok{ Ctx)}\OperatorTok{:} \BuiltInTok{Promise}\OperatorTok{\textless{}}\NormalTok{\{ refundableAmount}\OperatorTok{:} \DataTypeTok{bigint}\OperatorTok{;}\NormalTok{ refundPolicyRef}\OperatorTok{:} \DataTypeTok{string}\OperatorTok{;}
\NormalTok{      quoteExpiresAt}\OperatorTok{:} \DataTypeTok{number}\OperatorTok{;}\NormalTok{ engineResponse}\OperatorTok{:} \DataTypeTok{unknown}\NormalTok{ \}}\OperatorTok{\textgreater{};}      \CommentTok{// refund quote}
\NormalTok{\}}
\end{Highlighting}
\end{Shaded}

Every call carries a \texttt{Ctx} with the identifiers above so the connector can make the engine call idempotent on \texttt{(engineRef,\ chargeId,\ idempotencyKey)}. \texttt{createHold} MUST return \texttt{engineExpiry} when the engine can hold before authorization and \texttt{null} otherwise, in which case the operator applies the class's \texttt{policyExpiry} and the \texttt{holdBeforeAuth} path (Section 4.4). \texttt{commit} MUST refuse, without calling the engine, if the current time is at or past the charge's \texttt{issueDeadline}. \textbf{Timeout rule.} If \texttt{commit} times out or returns an indeterminate error, the operator MUST NOT retry it; it MUST call \texttt{getHold} and act on the engine's reported state (\texttt{committed} → produce the receipt and capture; \texttt{held} → retry \texttt{commit} if still before \texttt{issueDeadline}; \texttt{cancelled}/\texttt{expired} → void; \texttt{unknown} → keep polling until \(t_{\text{issue}}\), then void). A retried \texttt{commit} without a preceding \texttt{getHold} risks duplicate orders, double bookings or double ticket issuance and is non-conformant. The connector holds the attestor key for its class; the operator MUST register that key on-chain via \texttt{setAttestor} for the sellers the connector serves.

\textbf{E.3 Seller registry entry} (fields marked \emph{pub} are served by \texttt{/registry/sellers}):

\begin{Shaded}
\begin{Highlighting}[]
\FunctionTok{\{}
  \DataTypeTok{"sellerId"}\FunctionTok{:} \StringTok{"0x\textless{}bytes32\textgreater{}"}\FunctionTok{,}                    \ErrorTok{//} \ErrorTok{pub}
  \DataTypeTok{"payoutVault"}\FunctionTok{:} \StringTok{"0x\textless{}address\textgreater{}"}\FunctionTok{,}                 \ErrorTok{//} \ErrorTok{pub}
  \DataTypeTok{"reserveVault"}\FunctionTok{:} \StringTok{"0x\textless{}address\textgreater{}"}\FunctionTok{,}
  \DataTypeTok{"acceptedTokens"}\FunctionTok{:} \OtherTok{[}\StringTok{"0x\textless{}usdc\textgreater{}"}\OtherTok{]}\FunctionTok{,}               \ErrorTok{//} \ErrorTok{pub}
  \DataTypeTok{"classes"}\FunctionTok{:} \OtherTok{[}\StringTok{"order.shipped"}\OtherTok{,} \StringTok{"travel.air"}\OtherTok{]}\FunctionTok{,}   \ErrorTok{//} \ErrorTok{pub}
  \DataTypeTok{"delegations"}\FunctionTok{:} \OtherTok{[}\FunctionTok{\{} \DataTypeTok{"signer"}\FunctionTok{:} \StringTok{"0x\textless{}connector key\textgreater{}"}\FunctionTok{,}
      \DataTypeTok{"scope"}\FunctionTok{:} \FunctionTok{\{} \DataTypeTok{"classes"}\FunctionTok{:} \OtherTok{[}\StringTok{"*"}\OtherTok{]}\FunctionTok{,} \DataTypeTok{"maxAmount"}\FunctionTok{:} \StringTok{"5000000000"}\FunctionTok{,}
                 \DataTypeTok{"validUntil"}\FunctionTok{:} \DecValTok{1790000000} \FunctionTok{\},} \DataTypeTok{"revoked"}\FunctionTok{:} \KeywordTok{false} \FunctionTok{\}}\OtherTok{]}\FunctionTok{,}   \ErrorTok{//} \ErrorTok{pub}
  \DataTypeTok{"attestors"}\FunctionTok{:} \OtherTok{[}\StringTok{"0x\textless{}connector attestor key\textgreater{}"}\OtherTok{]}\FunctionTok{,}  \ErrorTok{//} \ErrorTok{pub}
  \DataTypeTok{"riskTier"}\FunctionTok{:} \StringTok{"verified"}\FunctionTok{,}                       \ErrorTok{//} \ErrorTok{pub}
  \DataTypeTok{"feeBps"}\FunctionTok{:} \DecValTok{150}\FunctionTok{,} \DataTypeTok{"distributorShareBps"}\FunctionTok{:} \DecValTok{6000}\FunctionTok{,}
  \DataTypeTok{"reserveBps"}\FunctionTok{:} \DecValTok{500}\FunctionTok{,} \DataTypeTok{"settlementDelaySeconds"}\FunctionTok{:} \DecValTok{604800}\FunctionTok{,}
  \DataTypeTok{"exposureCap"}\FunctionTok{:} \StringTok{"250000000000"}\FunctionTok{,} \DataTypeTok{"perChargeCap"}\FunctionTok{:} \StringTok{"5000000000"}\FunctionTok{,}
  \DataTypeTok{"dailyCap"}\FunctionTok{:} \StringTok{"100000000000"}\FunctionTok{,}
  \DataTypeTok{"settlementMode"}\FunctionTok{:} \StringTok{"escrow"}
\FunctionTok{\}}
\end{Highlighting}
\end{Shaded}

The on-chain vault MUST hold, at minimum, \texttt{payoutVault}, \texttt{reserveVault}, the delegation set, the attestor set, \texttt{feeBps}, \texttt{reserveBps} and \texttt{settlementDelaySeconds}; the rest MAY be off-chain.

\textbf{E.4 Fulfilment-class registry entry:}

\begin{Shaded}
\begin{Highlighting}[]
\FunctionTok{\{}
  \DataTypeTok{"classId"}\FunctionTok{:} \StringTok{"service.appointment"}\FunctionTok{,}
  \DataTypeTok{"engineOps"}\FunctionTok{:} \FunctionTok{\{} \DataTypeTok{"hold"}\FunctionTok{:} \StringTok{"slot.tentative"}\FunctionTok{,} \DataTypeTok{"commit"}\FunctionTok{:} \StringTok{"slot.confirm"}\FunctionTok{,}
                 \DataTypeTok{"cancel"}\FunctionTok{:} \StringTok{"slot.cancel"}\FunctionTok{,} \DataTypeTok{"refundQuote"}\FunctionTok{:} \StringTok{"policy.noticePeriod"} \FunctionTok{\},}
  \DataTypeTok{"holdLimitSource"}\FunctionTok{:} \StringTok{"engine"}\FunctionTok{,}              \ErrorTok{//} \ErrorTok{engine} \ErrorTok{|} \ErrorTok{policy} \ErrorTok{|} \ErrorTok{invoiceDueDate}
  \DataTypeTok{"holdBeforeAuth"}\FunctionTok{:} \KeywordTok{true}\FunctionTok{,}
  \DataTypeTok{"policyExpirySeconds"}\FunctionTok{:} \KeywordTok{null}\FunctionTok{,}
  \DataTypeTok{"minHoldSeconds"}\FunctionTok{:} \DecValTok{900}\FunctionTok{,}
  \DataTypeTok{"escrowMode"}\FunctionTok{:} \KeywordTok{true}\FunctionTok{,}                        \ErrorTok{//} \ErrorTok{false} \ErrorTok{=\textgreater{}} \ErrorTok{charge()} \ErrorTok{only}
  \DataTypeTok{"defaultRung"}\FunctionTok{:} \StringTok{"C"}\FunctionTok{,}
  \DataTypeTok{"fulfilmentRefKind"}\FunctionTok{:} \StringTok{"completionId"}\FunctionTok{,}
  \DataTypeTok{"refundRuleSource"}\FunctionTok{:} \StringTok{"engine"}\FunctionTok{,}
  \DataTypeTok{"challengeWindowSeconds"}\FunctionTok{:} \DecValTok{86400}\FunctionTok{,}
  \DataTypeTok{"refundWindowSeconds"}\FunctionTok{:} \DecValTok{2592000}
\FunctionTok{\}}
\end{Highlighting}
\end{Shaded}

\textbf{E.5 Arbiter interface.} An arbiter MUST expose \texttt{open(case)}, \texttt{submitEvidence(caseId,\ party,\ evidence)} and \texttt{rule(caseId)\ -\textgreater{}\ \{\ outcome:\ "uphold"\ \textbar{}\ "refund",\ amount\ \}}. A \texttt{refund} ruling is executed by the operator through the liquidity order and recorded with \texttt{source} on-chain. The arbiter MAY be a contract (rung D/S with on-chain evidence) or an off-chain service whose rulings are signed; the Offer's \texttt{verificationRung} determines which is acceptable.

\textbf{E.6 Webhook payloads.} Delivered as \texttt{POST} with an HMAC signature header over the body, retried with backoff for 24 h, and MUST be idempotent on \texttt{eventId}.

\begin{Shaded}
\begin{Highlighting}[]
\FunctionTok{\{} \DataTypeTok{"eventId"}\FunctionTok{:} \StringTok{"uuid"}\FunctionTok{,} \DataTypeTok{"aspVersion"}\FunctionTok{:} \StringTok{"1"}\FunctionTok{,} \DataTypeTok{"type"}\FunctionTok{:} \StringTok{"charge.captured"}\FunctionTok{,}
  \DataTypeTok{"chargeId"}\FunctionTok{:} \StringTok{"0x…"}\FunctionTok{,} \DataTypeTok{"sellerId"}\FunctionTok{:} \StringTok{"0x…"}\FunctionTok{,} \DataTypeTok{"occurredAt"}\FunctionTok{:} \DecValTok{1790000000}\FunctionTok{,}
  \DataTypeTok{"txHash"}\FunctionTok{:} \StringTok{"0x…"}\FunctionTok{,}
  \DataTypeTok{"data"}\FunctionTok{:} \FunctionTok{\{} \DataTypeTok{"receiptHash"}\FunctionTok{:} \StringTok{"0x…"}\FunctionTok{,} \DataTypeTok{"toSeller"}\FunctionTok{:} \StringTok{"504710000"}\FunctionTok{,}
            \DataTypeTok{"fee"}\FunctionTok{:} \StringTok{"7690000"}\FunctionTok{,} \DataTypeTok{"reserve"}\FunctionTok{:} \StringTok{"0"} \FunctionTok{\}} \FunctionTok{\}}
\end{Highlighting}
\end{Shaded}

Event types: \texttt{charge.authorized}, \texttt{charge.captured}, \texttt{charge.voided}, \texttt{charge.reclaimed}, \texttt{charge.refunded} (with \texttt{cumulative} and \texttt{final:\ bool}), \texttt{charge.challenged}, \texttt{challenge.ruled}, \texttt{charge.late\_capture\_loss} (operator-internal), \texttt{hold.released}.

\textbf{E.7 Time and identity conventions.} All timestamps are unix seconds UTC. Chain identifiers are CAIP-2. \texttt{sellerId}, \texttt{chargeId}, \texttt{offerId} and \texttt{receiptHash} are \texttt{bytes32} rendered as 0x-prefixed hex. Amounts are decimal strings in the token's smallest unit. The connector MUST convert engine-local time to UTC unix seconds and MUST apply the class's \(\delta\) when doing so. \textbf{Clock health.} The operator MUST monitor its clock against the chain's latest block timestamp and at least one trusted time source, and MUST stop issuing Offers while observed drift exceeds the configured \(\delta\). Clock drift is a safety input to H1 and is treated as infrastructure health, like RPC availability.

\textbf{E.8 Relayer requirements.} The relayer is part of the safety-critical path because \(T_s\) and \(T_i\) are its properties. It MUST record, per transaction, \texttt{submittedAt}, \texttt{includedAt}, \texttt{finalizedAt}, nonce, RPC endpoint, retries, replacements and fee parameters; MUST use at least two independent RPC endpoints; MUST implement nonce management and fee-bump replacement; and MUST alert when observed \(T_s + T_i\) over a rolling window exceeds a configured fraction of \(M\). The evaluation plan (Section 9) reports the empirical \(T_s + T_i\) distribution from these records.

\textbf{E.9 Authority of record.} Three stores hold state and the specification fixes which is authoritative for what: the chain for payment state; the fulfilment engine for fulfilment state; the operator database for orchestration and derived state only. Any disagreement is resolved by reconciliation toward the authoritative source, never by editing it. The operator MUST run reconciliation over every non-terminal charge (the reconciliation matrix is given in the ASP-Lite Implementation Specification and summarised in Section 4.11).

\hypertarget{appendix-f-conformance-levels-and-versioning}{%
\section{Conformance levels and versioning}\label{appendix-f-conformance-levels-and-versioning}}

An implementation MAY claim one of three levels.

\begin{longtable}[]{@{}
  >{\raggedright\arraybackslash}p{(\columnwidth - 2\tabcolsep) * \real{0.1000}}
  >{\raggedright\arraybackslash}p{(\columnwidth - 2\tabcolsep) * \real{0.9000}}@{}}
\toprule\noalign{}
\begin{minipage}[b]{\linewidth}\raggedright
Level
\end{minipage} & \begin{minipage}[b]{\linewidth}\raggedright
MUST support
\end{minipage} \\
\midrule\noalign{}
\endhead
\bottomrule\noalign{}
\endlastfoot
\textbf{ASP-Lite} & One token; escrow mode; rung C; Appendix D vault; Appendix C scheme; E.1 routes \texttt{/quotes}, \texttt{/charges}, \texttt{/charges/\{id\}}, \texttt{/charges/\{id\}/receipt}, \texttt{/charges/\{id\}/cancel}, \texttt{/charges/\{id\}/refund}; one connector with the E.2 timeout rule; delegated Offer signing with version-based revocation; a single attestor grant; exposure reservation (Section 4.6); E.7 clock health and E.8 relayer records; Appendix B cases 1--17. \\
\textbf{ASP-Core} & ASP-Lite plus: multiple tokens under Invariant I1; distributor share (Proposition 2); all four rungs with an E.5 arbiter; the \texttt{holdBeforeAuth} path; E.6 webhooks; exposure controls with the Proposition 3 ceiling. \\
\textbf{ASP-Full} & ASP-Core plus: instant-reserve mode; on-chain hooks under P5; ERC-8004 identity; partial capture. \\
\end{longtable}

Versioning: \texttt{aspVersion} is a single integer carried in every message. Additive changes (new optional fields, event types, error codes) do not bump it; any change to the EIP-712 struct layouts, the vault interface, or the meaning of an existing field MUST bump it, and a vault MUST reject authorizations whose EIP-712 domain \texttt{version} it does not implement. Fulfilment-class registries and connectors are versioned independently of the protocol.

Conformance is demonstrated by running the executable suite that accompanies the reference implementation (\texttt{asp-conformance}, Section 9), which reports the number of Appendix B cases passed at the claimed level.

Interoperability claim: two implementations at the same level that each pass Appendix B, expose Appendix D, and speak Appendix C SHOULD be able to exchange a buyer agent, a seller, or a connector without modification. Demonstrating this with two independent ASP-Lite implementations is part of the evaluation plan (Section 9).

  \bibliography{refs}

@misc{x402,
  author = {Reppel, Erik and Dalal, Nemil and Kim, Dan},
  title = {Introducing x402: a new standard for internet-native payments},
  howpublished = {Coinbase Developer Platform},
  year = {2025},
  url = {https://www.coinbase.com/developer-platform/discover/launches/x402}
}

@misc{x402schemes,
  author = {{x402 Foundation}},
  title = {Payment Schemes: exact, upto, and batch-settlement},
  howpublished = {x402 documentation},
  year = {2026},
  url = {https://docs.x402.org/schemes/overview}
}

@misc{x402lf,
  author = {{Linux Foundation}},
  title = {x402 transfers to the Linux Foundation},
  year = {2026},
  month = jul,
  note = {Press coverage: https://xenospectrum.com/en/linux-foundation-x402-agent-payments/}
}

@misc{cloudflare402,
  author = {{Cloudflare}},
  title = {Launching the x402 Foundation with Coinbase, and support for x402 transactions},
  howpublished = {Cloudflare Blog},
  year = {2025},
  month = sep,
  url = {https://blog.cloudflare.com/x402/}
}

@misc{cpp,
  author = {{Coinbase} and {Shopify}},
  title = {Commerce Payments Protocol: Onchain authorization and capture for trust-minimized commerce},
  howpublished = {GitHub repository base/commerce-payments},
  year = {2025},
  url = {https://github.com/base/commerce-payments}
}

@misc{cppblog,
  author = {{Shopify Engineering}},
  title = {Engineering the Commerce Payments Protocol},
  year = {2025},
  month = jun,
  url = {https://shopify.engineering/commerce-payments-protocol}
}

@misc{x402r,
  author = {{x402r}},
  title = {x402r Refund Protocol: Escrow-backed refunds for x402 payments},
  year = {2026},
  url = {https://www.x402r.org/}
}

@misc{ap2,
  author = {Parikh, Stavan and Surapaneni, Rao},
  title = {Powering AI commerce with the new Agent Payments Protocol (AP2)},
  howpublished = {Google Cloud Blog},
  year = {2025},
  url = {https://cloud.google.com/blog/products/ai-machine-learning/announcing-agents-to-payments-ap2-protocol}
}

@misc{acp,
  author = {{Stripe} and {OpenAI}},
  title = {Agentic Commerce Protocol},
  year = {2025},
  url = {https://www.agenticcommerce.dev/}
}

@misc{erc3009,
  author = {Lecoq, Peter and Fischer, Yoav},
  title = {EIP-3009: Transfer With Authorization},
  howpublished = {Ethereum Improvement Proposals},
  year = {2020},
  url = {https://eips.ethereum.org/EIPS/eip-3009}
}

@misc{eip2612,
  author = {Lecoq, Peter},
  title = {EIP-2612: Permit Extension for EIP-20 Signed Approvals},
  howpublished = {Ethereum Improvement Proposals},
  year = {2020},
  url = {https://eips.ethereum.org/EIPS/eip-2612}
}

@misc{eip712,
  author = {Bloemen, Remco and Logvinov, Leonid and Evans, Jacob},
  title = {EIP-712: Typed structured data hashing and signing},
  howpublished = {Ethereum Improvement Proposals},
  year = {2017},
  url = {https://eips.ethereum.org/EIPS/eip-712}
}

@misc{erc4337,
  author = {Buterin, Vitalik and Weiss, Yoav and Tirosh, Dror and Nacson, Shahaf and Forshtat, Alex and Gazso, Kristof and Hess, Tjaden},
  title = {ERC-4337: Account Abstraction Using Alt Mempool},
  howpublished = {Ethereum Improvement Proposals},
  year = {2021},
  url = {https://eips.ethereum.org/EIPS/eip-4337}
}

@misc{erc8004,
  author = {{ERC-8004 Authors}},
  title = {ERC-8004: Trustless Agents},
  howpublished = {Ethereum Improvement Proposals},
  year = {2025},
  url = {https://eips.ethereum.org/EIPS/eip-8004}
}

@misc{permit2,
  author = {{Uniswap Labs}},
  title = {Permit2},
  howpublished = {GitHub repository Uniswap/permit2},
  year = {2022},
  url = {https://github.com/Uniswap/permit2}
}

@article{sokA2A,
  author = {Zhang, Yuanzhe and others},
  title = {{SoK}: Blockchain Agent-to-Agent Payments},
  journal = {arXiv preprint arXiv:2604.03733},
  year = {2026},
  note = {Author list to be verified against the arXiv record}
}

@article{sokAgenticCommerce,
  author = {Mao, Qian'ang and Wang, Jiaxin and Liu, Ya and Zhu, Li and Ma, Cong and Yan, Jiaqi},
  title = {{SoK}: Security of Autonomous {LLM} Agents in Agentic Commerce},
  journal = {arXiv preprint arXiv:2604.15367},
  year = {2026}
}

@article{a402,
  author = {Li, Yue and Wang, Lei and Wang, Kaixuan and Yang, Zhiqiang and Wang, Ke and Guan, Zhi and Gao, Jianbo},
  title = {A402: Binding Cryptocurrency Payments to Service Execution for Agentic Commerce},
  journal = {arXiv preprint arXiv:2603.01179},
  year = {2026}
}

@article{freeriding,
  author = {{Anonymous}},
  title = {Free-Riding the Agentic Web: A Systematic Security Analysis of x402 Payments},
  journal = {arXiv preprint arXiv:2605.30998},
  year = {2026},
  note = {Author list to be verified against the arXiv record}
}

@article{compliance,
  author = {See, Kenneth and Tan, Xue Wen},
  title = {Compliance-Aware Agentic Payments on Stablecoin Rails},
  journal = {arXiv preprint arXiv:2605.00071},
  year = {2026}
}

@article{erc8004study,
  author = {Xiong, Xiaofan and Li, Zhuo and Wei, Wenjie and Wang, Qin and Knottenbelt, William and Wang, Zhipeng},
  title = {Can Trustless Agents Be Trusted? An Empirical Study of the {ERC-8004} Decentralized {AI} Agent Ecosystem},
  journal = {arXiv preprint arXiv:2606.26028},
  year = {2026}
}

@article{ap2replay,
  author = {Lan, Qianlong and Kaul, Anuj and Jones, Shaun and Westrum, Stephanie},
  title = {Zero-Trust Runtime Verification for Agentic Payment Protocols: Mitigating Replay and Context-Binding Failures in {AP2}},
  journal = {arXiv preprint arXiv:2602.06345},
  year = {2026}
}

@article{vaziry,
  author = {Vaziry, Awid and Rodriguez Garzon, Sandro and K{\"u}pper, Axel},
  title = {Towards Multi-Agent Economies: Enhancing the {A2A} Protocol with Ledger-Anchored Identities and x402 Micropayments for {AI} Agents},
  journal = {arXiv preprint arXiv:2507.19550},
  year = {2025}
}

@article{aesp,
  author = {Wang, J. S.},
  title = {{AESP}: A Human-Sovereign Economic Protocol for {AI} Agents with Privacy-Preserving Settlement},
  journal = {arXiv preprint arXiv:2603.00318},
  year = {2026}
}

@article{alqithami,
  author = {Alqithami, Saad},
  title = {Autonomous Agents on Blockchains: Standards, Execution Models, and Trust Boundaries},
  journal = {arXiv preprint arXiv:2601.04583},
  year = {2026}
}

@inproceedings{pang2002,
  author = {Pang, Xiaolin and Tan, Kian-Lee and Wang, Yong and Ren, Jian},
  title = {A Secure Agent-Mediated Payment Protocol},
  booktitle = {Information and Communications Security (ICICS)},
  publisher = {Springer},
  pages = {422--433},
  year = {2002}
}

@misc{ndc,
  author = {{IATA}},
  title = {New Distribution Capability (NDC) Standard},
  howpublished = {International Air Transport Association},
  year = {2024},
  url = {https://www.iata.org/en/programs/airline-distribution/retailing/ndc/}
}

@misc{xdpos2,
  author = {{XDC Network}},
  title = {XDPoS 2.0: Deterministic Finality for the XDC Network},
  howpublished = {XDC Network documentation},
  year = {2023},
  url = {https://docs.xdc.network/}
}

@misc{tla,
  author = {Lamport, Leslie},
  title = {Specifying Systems: The {TLA+} Language and Tools for Hardware and Software Engineers},
  publisher = {Addison-Wesley},
  year = {2002}
}

@misc{rfc2119,
  author = {Bradner, Scott},
  title = {Key words for use in {RFCs} to Indicate Requirement Levels},
  howpublished = {RFC 2119, IETF},
  year = {1997},
  url = {https://www.rfc-editor.org/rfc/rfc2119}
}

@misc{caip2,
  author = {{Chain Agnostic Standards Alliance}},
  title = {CAIP-2: Blockchain ID Specification},
  year = {2019},
  url = {https://github.com/ChainAgnostic/CAIPs/blob/main/CAIPs/caip-2.md}
}

@misc{openapi,
  author = {{OpenAPI Initiative}},
  title = {OpenAPI Specification v3.1},
  year = {2021},
  url = {https://spec.openapis.org/oas/v3.1.0}
}

\end{document}